\documentclass[12pt]{article}
\usepackage{times}
\usepackage{graphicx}

\usepackage[style=science, maxbibnames=99]{biblatex}
\newenvironment{sciabstract}{%
\begin{quote} \bf}
{\end{quote}}

\title{Quantum behavior of a superconducting Duffing oscillator at the dissipative phase transition} 

\author
{Qi-Ming Chen$^{1,2,\ast}$, Michael Fischer$^{1,2}$, Yuki Nojiri$^{1,2}$,\\ 
Michael Renger$^{1,2}$, Edwar Xie$^{1,2}$, Matti Partanen$^{1}\dagger$,\\
Stefan Pogorzalek$^{1,2}\ddagger$, Kirill G. Fedorov$^{1,2}$, Achim Marx$^{1}$,\\
Frank Deppe$^{1,2,3,\ast}\ddagger$, Rudolf Gross $^{1,2,3,\ast}$\\
\\
\normalsize{$^{1}$Walther-Mei{\ss}ner-Institut, Bayerische Akademie der Wissenschaften; 85748 Garching, Germany}\\
\normalsize{$^{2}$Physik-Department, Technische Universit{\"a}t M{\"u}nchen; 85748 Garching, Germany}\\
\normalsize{$^{3}$Munich Center for Quantum Science and Technology (MCQST); 80799 Munich, Germany}\\
\normalsize{$^{\ast}$Corresponding author. E-mail: qiming.chen@wmi.badw.de (Q.C.);}\\
\normalsize{frank.deppe@wmi.badw.de (F.D.); rudolf.gross@wmi.badw.de (R.G.)}\\
\normalsize{Present addresses: $\dagger$IQM; 02150 Espoo, Finland; $\ddagger$IQM; 80335 M{\"u}nchen, Germany}
}

\date{}

\begin{document} 
\baselineskip24pt
\maketitle 

\begin{sciabstract}
Understanding the non-deterministic behavior of deterministic nonlinear systems has been an implicit dream since Lorenz named it the ``butterfly effect". A prominent example is the hysteresis and bistability of the Duffing oscillator, which in the classical description is attributed to the coexistence of two steady states in a double-well potential. However, this interpretation fails in the quantum-mechanical perspective, where a single unique steady state is allowed in the whole parameter space. Here, we measure the non-equilibrium dynamics of a superconducting Duffing oscillator and reconcile the classical and quantum descriptions in a unified picture of quantum metastability. We demonstrate that the two classically regarded steady states are in fact metastable states. They have a remarkably long lifetime in the classical hysteresis regime but must eventually relax into a single unique steady state allowed by quantum mechanics. By engineering the lifetime of the metastable states sufficiently large, we observe a first-order dissipative phase transition, which mimics a sudden change of the mean field in a $11$-site Bose-Hubbard lattice. We also reveal the two distinct phases of the transition by quantum state tomography, namely a coherent-state phase and a squeezed-state phase separated by a critical point. Our results reveal a smooth quantum state evolution behind a sudden dissipative phase transition, and they form an essential step towards understanding hysteresis and instability in non-equilibrium systems.
\end{sciabstract}

\section*{One-Sentence Summary}
Tomography of phase transition process reveals non-equilibrium quantum dynamics of the Duffing oscillator.

\section*{Main Text}
The Duffing oscillator is a simple but prototypical model in nonlinear physics, which describes a forced oscillation with cubic nonlinearity and linear viscous damping \cite{Nayfeh1995}. In a certain parameter regime, classical mechanics predicts a double-well potential that allows two steady states (SSs) at the same parameter setting \cite{Landau1976}. It gives rise to a hysteretic behavior where two different amplitudes of the forced oscillation are possible. Depending on whether the system is initially at rest or in strong oscillation, the oscillator spontaneously chooses one of the amplitudes when adiabatically tuning the parameters into the hysteresis regime. Thermal fluctuations may induce unpredictable jumps between the two potential wells and lead to the bistability of the oscillation amplitude. This classical behavior of the Duffing oscillator has been observed in a considerable number of experiments, for example, in superconducting quantum circuits \cite{Siddiqi2004,Siddiqi2005}. The underlying double-well potential model has been used to explain a variety of physical processes, such as optical bistability \cite{Gibbs1976,Rempe1991}, parametric amplification \cite{Yurke1984,Lin2014}, and self-oscillation \cite{Fajans2001,Murch2010}. However, it has been revealed by Drummond and Walls already in the 1980s that a fully-quantum treatment of the Duffing oscillator yields a single unique SS over the entire parameter space, such that it ``does not exhibit bistability or hysteresis" \cite{Drummond1980}. These two perspectives indicate fundamentally different behaviors of the Duffing oscillator. However, the seeming two classical SSs are still observed even in a typical quantum experiment setup \cite{Siddiqi2005,Naaman2008}. Important experimental milestones towards revealing the quantum behavior include the observation of intriguing phenomena in the transmission spectrum \cite{Mavrogordatos2017,Brookes2021}, decay rate \cite{Rodriguez2017,Brookes2021}, and second-order correlation function \cite{Fink2017} around a specific parameter setting in a continuous-wave measurement setup.

Here, we simulate the non-equilibrium quantum dynamics of the Duffing oscillator with an \textit{in-situ} tunable superconducting nonlinear resonator in a pulsed measurement setup. Our experimental results settle the seeming controversy between the classical and quantum properties of the Duffing oscillator. We demonstrate that the two classical SSs are in fact metastable states (MSs), which emerge when the low-lying eigenvalues of the Liouvillian superoperator are separated from the rest of its spectrum \cite{Macieszczak2016}. In the classical hysteresis regime, the MSs have a lifetime much longer than any other timescale in the system, but are not the exact SS solutions of the Schr{\"o}dinger equation. An exceptional case occurs when the system approaches the thermodynamic limit, which mimics the mean-field model of a driven-dissipative Bose-Hubbard lattice \cite{Casteels2017}. In this case, the MSs gain an increasingly long lifetime when approaching a critical point but, suddenly, cannot be properly defined at the exact point \cite{Minganti2018}. This non-analytical phenomenon is identified as a first-order dissipative phase transition (DPT), which originates from the interplay between a coherent drive and an incoherent dissipation in a driven-dissipative system.

The non-equilibrium quantum dynamics of the Duffing oscillator is described by the master equation in the Born-Markov approximation: $\partial_t \rho(t) = \mathcal{L}\rho(t)$, where the Liouvillian superoperator, $\mathcal{L}$, consists of the system Hamiltonian, $H/\hbar = \Delta a^{\dagger}a + U a^{\dagger}a^{\dagger}aa + \xi \left(a+a^{\dagger}\right)$, and the Lindblad superoperator, $\left(\gamma/2\right)\mathcal{D}\left[a\right]$. Besides, $a$ ($a^{\dagger}$) is the annihilation (creation) operator of the oscillating mode, $\Delta$ the detuning between the resonant frequency, $\omega_{\rm A}$, and the drive, $U$ the Kerr nonlinearity, and $\xi$ the driving strength. We define $\gamma$ as the total energy dissipation rate, which is approximately $3.85\,{\rm \mu s^{-1}}$ in the measured frequency range, and we neglect the relatively weak dephasing effect \cite{supplement}. When restricting our discussion to finite dimensions, the Liouvillian superoperator can be decomposed into Jordan blocks that lead to the formal solution: $\rho (t) = \sum_{n} \exp\left(\lambda_n t\right) \left(\sum_m c_{n,m}r_{n,m}\right)$ with $c_{n,m}={\rm tr}\left[ l_{n,m}\rho(0) \right]$. Here, $l_{n,m}$ and $r_{n,m}$ are the left and right eigenmatrices of $\mathcal{L}$, which correspond to the $n$th eigenvalue with geometric multiplicity $m=1,2,\cdots$. For convenience, we define $\delta_n=-{\rm Re}\left(\lambda_n\right)$ and sort the eigenvalues according to $\delta_{n} < \delta_{n+1}$. Under quite general conditions, there exists a single unique SS solution such that $\delta_0 = 0$, $\delta_1 > 0$ \cite{Albert2014}. Thus, the smallest nonzero eigenvalue, forming the Liouvillian gap $\delta_1$, determines the timescale the system requires to relax into the SS solution, and thereby results in a general exponential decay of an observable. However, if the Liouvillian gap is well separated from the rest of its spectrum, $\delta_1 \ll \delta_2$, the system may quickly relax onto the metastable manifold spanned by $r_{0,1}$ and $\left\{ r_{1,m} \right\}$ within a timescale of $1/\delta_2$, and stays almost invariant for a relatively large timescale, $1/\delta_1$, before starting a second relaxation into the unique SS solution \cite{Macieszczak2016}. The Liouvillian gap may even close at a critical point in the thermodynamic limit, where the eigenvalue \textit{zero} has a geometric multiplicity of two but an algebraic multiplicity of one. The SS thus must undergo a sudden change on the two sides of the critical point and result in a first-order DPT \cite{Minganti2018}. 

The system studied in our experiment is realized by embedding a weakly asymmetric superconducting quantum interference device (SQUID) in the middle of a coplanar waveguide resonator. By driving it with a coherent microwave field, we implement a Duffing oscillator in superconducting quantum circuits with tunable frequency and nonlinearity \cite{Leib2012, Fischer2021}, as shown in Fig.\,1A. In our experiment, the resonant frequency, $\omega_A/2\pi$, is varied between $6.80\,{\rm GHz}$ and $7.15\,{\rm GHz}$, corresponding to a tunable range of the nonlinearity from $U/2\pi=-295\,{\rm kHz}$ to $-58\,{\rm kHz}$. We modulate the radio-frequency drive by three different pulse shapes, which prepare the system in one of the two potential wells or in the SS at the initial time. Then, we switch the driving strength to $\xi$, and trigger a short measurement of the transmitted or reflected microwave signal after a time delay of $\tau$. In each repetition, the measurement lasts for only $16\,{\rm ns}$ to capture the transient non-equilibrium dynamics of the system. We repeat this procedure for $10^{6}$ -- $10^{9}$ times depending on the required accuracy, and concatenate the results in a long trace for extracting the quadrature histogram of the transient field. Eventually, we obtain the quasi-distribution functions of the intra-resonator field for different initial states and also different control parameters, $\Delta$, $\xi$, and $\tau$ \cite{supplement}. 

In our experiment, we first tune the nonlinearity to $U/2\pi=-132\,{\rm kHz}$ and drive the system with a varying strength, $\xi$, and detuning, $\Delta$. The measurement is delayed by $\tau=3.25\,{\rm \mu s}$, which is more than $10$ times longer than the free-relaxation time of the resonator, $1/\gamma$. When the system is initially prepared in one of the two potential wells, the absolute mean field, $\left|\langle a \rangle\right|$, and the photon number, $\langle a^{\dagger}a \rangle$, show an abrupt change at either of the two boundaries of the classical hysteresis regime, as shown in Fig.\,1B. Within this regime, the measured values are also different for the same parameter setting, which correspond to the two possible oscillation amplitudes of the Duffing oscillator, i.e., the two classical SSs in a double-well potential. However, the transition occurs inside the regime when applying a constant driving field, which corresponds to an infinitely large measurement delay in either of the two former cases. Classically, this is explained by the presence of thermal fluctuations that induce random jumps between the two potential wells and wash out the dependence on the initial-state for large $\tau$. However, this interpretation fails in our experimental situation where the thermal noise at the $30\,{\rm mK}$ base temperature is much smaller than half a photon, and thus not likely to cause a noticeable transition between the two potential wells. 

Indeed, quantum fluctuations play a significant role in the hysteresis regime, as shown in Fig.\,1C for a fixed detuning frequency, $\Delta/2\pi=2.36\,{\rm MHz}$. A clear dip in the $|\langle a \rangle|$ curve is observed during the transition process, which is predicted as a result of out-of-phase quantum fluctuations for the unique SS of the Duffing oscillator \cite{Drummond1980,Mavrogordatos2017,Brookes2021}. By comparison, $\langle a^{\dagger}a \rangle$ is a monotonic function of $\xi$ since it is insensitive to the phase of quantum fluctuations. Moreover, the second-order correlation function, $g^{(2)}(0)$, is strongly peaked around the transition point and approaches unity for large $\xi$. This is a typical signature of a first-order DPT, resulting from the enhanced quantum fluctuations in the SS \cite{Drummond1980,Fink2017}. Remarkably, we observe these quantum-mechanical signatures in company with the classical hysteretic behavior, indicating that the system may have not reached the SS even for $\tau > 10/\gamma$. This new perspective on the $40$-year-old mystery suggests that the two classically regarded SSs may be interpreted as MSs with a remarkably long lifetime in the hysteresis regime. The specific MS, in which the systems is staying, is determined by the distance between the initial state and the two MSs. This leads to the seemingly classical behavior of hysteresis in Fig.\,1B.

According to the theory of quantum metastability, the MSs exist only in the time window $1/\delta_2 \ll \tau \leq 1/\delta_1$ and should eventually relax into the single unique SS for $\tau \gg 1/\delta_1$ \cite{Macieszczak2016}. To verify this prediction, we then fix the detuning frequency at $\Delta/2\pi=2.01\,{\rm MHz}$ and measure the complex reflection coefficient, $S_{22}$, for varying driving strength, $\xi$, and measurement delay, $\tau$. In these measurements, the nonlinearity is fixed at $U/2\pi=-71\,{\rm kHz}$ \cite{supplement}. In Fig.\,2A, we plot the reflection coefficients, corresponding to the two MSs, in the complex plane. The two MS branches form a closed loop for each fixed $\tau$, manifesting the classical signature of hysteresis around $\xi^{*}/2\pi=1.51\,{\rm MHz}$. However, different from the classical interpretation the loop exists within a decreasing parameter range when increasing $\tau$. It is expected to close for $\tau > 55\,{\rm \mu s}$, where the two MS branches converge to the single unique SS allowed by quantum mechanics \cite{supplement}. This observation provides a clear evidence for the quantum description of the Duffing oscillator. It indicates that the hysteresis observed in the so-called quantum regime is the measurement outcome on two different MSs, while the system should eventually converge to a single unique SS in the long-time limit. 

To quantify this convergence, we calculate the loop area, $A$, for different $\tau$, as shown in Fig.\,2B. Fitting the data by an exponential decay, $A\propto \exp (-\eta \tau)$, we obtain two distinctively different decay rates $\eta_1 = 0.74\,{\rm \mu s^{-1}}$ and $\eta_2 = 0.04\,{\rm \mu s^{-1}}$ at small and large $\tau$, respectively. This two-stage relaxation process is qualitatively different from the classical prediction \cite{Jung1990}, but can be well understood from the quantum theory of Liouvillian spectrum \cite{Casteels2016,Rodriguez2017}. Fig.\,2C shows the fitted Liouvillian gap, $\delta_{1}$, as a function of the driving strength, $\xi$. The gap is approximately $3.79\,{\rm \mu s^{-1}}$ for both small and large $\xi$, what agrees well with the free energy decay rate $\gamma$. However, $\delta_1$ decreases by more than two orders of magnitude when approaching the critical point, $\xi^{*}$, and reaches a minimum value of $0.02\,{\rm \mu s^{-1}}$ at $\xi^{*}$. This observation indicates a critical slowing down of the system dynamics around the critical point. This is another signature of a first-order DPT \cite{Brookes2021}. For a sufficiently small $\tau$, the decay rate of the loop area, $\eta_1$, is determined by the average value of the Liouvillian gap over the hysteresis regime, that is, $1.22\,{\rm \mu s^{-1}}$. However, for $\tau \rightarrow \infty$ the decay rate $\eta_2$ is dominated by the minimum gap. In the time window between the two extreme cases, the decay rate decreases monotonically with $\tau$ and connects the two extremes. This is in quantitative agreement with the observed two-stage relaxation rates in Fig.\,2B.

It is then natural to ask whether the Liouvillian gap can be closed at a particular parameter setting, where the system dynamics becomes infinitely slow and the two MSs become also SSs. However, this perception is in conflict with the uniqueness of the SS solution for the Duffing oscillator \cite{Drummond1980}. Nevertheless, multiple SSs can exist in the driven-dissipative Bose-Hubbard model, where an infinite number of identical Duffing oscillators are coupled to each other and form a lattice. A bridge between the mean field description of an $N$-site Bose-Hubbard lattice and a single Duffing oscillator may be constructed by rescaling the nonlinearity and driving strength of the later as $U\rightarrow U_0/N$ and $\xi\rightarrow \sqrt{N}\xi_0$ \cite{Casteels2017}. A thermodynamic limit of the Duffing oscillator is achieved at $N\rightarrow\infty$, where the Liouvillian gap is closed at a rescaled critical point, $\xi_0^{*}$, and results in a first-order DPT. 

In our experiment, we \textit{in-situ} tune the scaling factor from approximately $N=2$ to $11$, and measure the average photon number of the SS for varying driving strength \cite{supplement}. Here, we define $U_0 \equiv -\gamma$ for $N=1$ and fix the detuning at $\Delta=3\gamma$ without loss of generality \cite{Casteels2017}. As shown in Fig.\,3, the transition happens at the same rescaled critical driving strength, $\xi_{0}^{*}/2\pi=0.58\,{\rm MHz}$, for different $N$, and the photon density also saturates at a similar value of around $\langle a^{\dagger}a \rangle/N=3.1$. However, the transition becomes increasingly sharp with increasing $N$. This observed tendency indicates a diverging lifetime of the MSs at the critical point in the thermodynamic limit, and thus a sudden change of the SS on the two sides of $\xi_{0}^{*}$, called the first-order DPT \cite{Minganti2018}. The Duffing oscillator thus explores two separate phases around the critical point.  

To understand the underlying physical process of DPT, we reconstruct the Wigner quasi-distribution function of the intra-resonator field according to the first two orders of signal moments, $\langle a \rangle$, $\langle a^{\dagger}a \rangle$, and $\langle a^2 \rangle$, as shown in Fig.\,4. Here, we operate the SQUID close to its sweet spot where the dephasing rate is sufficiently small \cite{supplement}, as indicated by the good agreement between theory and experiment in Fig.\,3. The reconstructed SS is approximately an either coherent or squeezed state in one of the two phases \cite{Bajer2004}, where the field mean coincides with the two classical SS solutions \cite{Casteels2017}. In each individual phase, the SS remains almost invariant with respect to the rescaled driving strength. However, the system undergoes a drastic change in a relatively small range around the critical point, $0.52\,{\rm MHz} \leq \xi_{0}/2\pi \leq 0.64\,{\rm MHz}$. This results in the rapid photon number transition in Fig.\,3. In this regime, the Wigner function consists of two separate parts in phase space, corresponding to the two SSs in the two individual phases \cite{Vogel1989,Kheruntsyan1999}. The probability of staying in the coherent-state phase changes continuously into that of being in the squeezed-state phase with increasing $\xi_{0}$. Ideally, it reaches a equiprobable mixture of the two phases at the exact critical point, $\xi_{0}^{*}/2\pi=0.58\,{\rm MHz}$ \cite{Minganti2018}. One can thus understand the dip of $\left|\langle a \rangle\right|$ and the peak of $g^{(2)}(0)$ in Fig.\,1C as a result of the coherent interference between the two phases. With the increase of $N$, the photon number diverges and the system behaves more classically. The SS thus must jump at $\xi_{0}^{*}$ in the thermodynamic limit, because only one potential well can be occupied at the same time in a classical system. This observation explains the increasingly sharp step of $\langle a^{\dagger}a \rangle$ with increasing $N$ in Fig.\,3, and reveals the origin of the first-order DPT.

The quantum behavior of the Duffing oscillator promotes the view that the extensively observed hysteresis and bistability originate from a non-classical SS around a critical point. The SS consists of two separate parts in phase space, which correspond to the two phases of the system and are MSs with a remarkably long lifetime. Their lifetime diverges when approaching the thermodynamic limit and leads to a first-order DPT. The tunable superconducting nonlinear resonator is a versatile building block for quantum simulation \cite{Raftery2014,Fitzpatrick2017,Ma2019}, and the pulsed heterodyne measurement enables the tomography of a non-equilibrium process. We therefore expect these methods to serve as a stepping stone for simulating strongly correlated bosons in the driven-dissipative regime \cite{Carusotto2013} and for unveiling the mystery of the ``butterfly effect" from a quantum-mechanical perspective. 


\printbibliography

\section*{Acknowledgments}
We thank P. Zapletal and M. J. Hartmann for insightful discussions of dephasing effect.
\paragraph*{Funding:}
German Research Foundation via Germany's Excellence Strategy (EXC-2111-390814868).
Elite Network of Bavaria through the program ExQM.
European Union via the Quantum Flagship project QMiCS (No.\,820505).
German Federal Ministry of Education and Research via the project QuaRaTe (No.\,13N15380).

\paragraph*{Author contributions:}
Q.C. carried out theoretical calculations, performed the experiment, and analyzed data. 
M.F. fabricated the sample and contributed to system characterization.
Y.N., M.R., and K.F. contributed to microwave techniques.
E.X. and A.M. contributed to cryogenic techniques.
M.P. and F.D. contributed to data analysis. 
S.P. contributed to FPGA programming.
Q.C., F.D., and R.G. wrote the manuscript with input from all authors. 
R.G. supervised the project.

\paragraph*{Competing interests:}
The authors declare no competing interests.

\paragraph*{Data and materials availability:}
The raw data and analysis code are available from the corresponding author (R.G.) on reasonable request.
 
\section*{Supplementary Materials}
Supplementary Text\\
Figs. S1 to S12\\
Table S1\\
References ($33$ -- $58$)

\newpage
\begin{figure}
  \centering
  \includegraphics[width=7.3in]{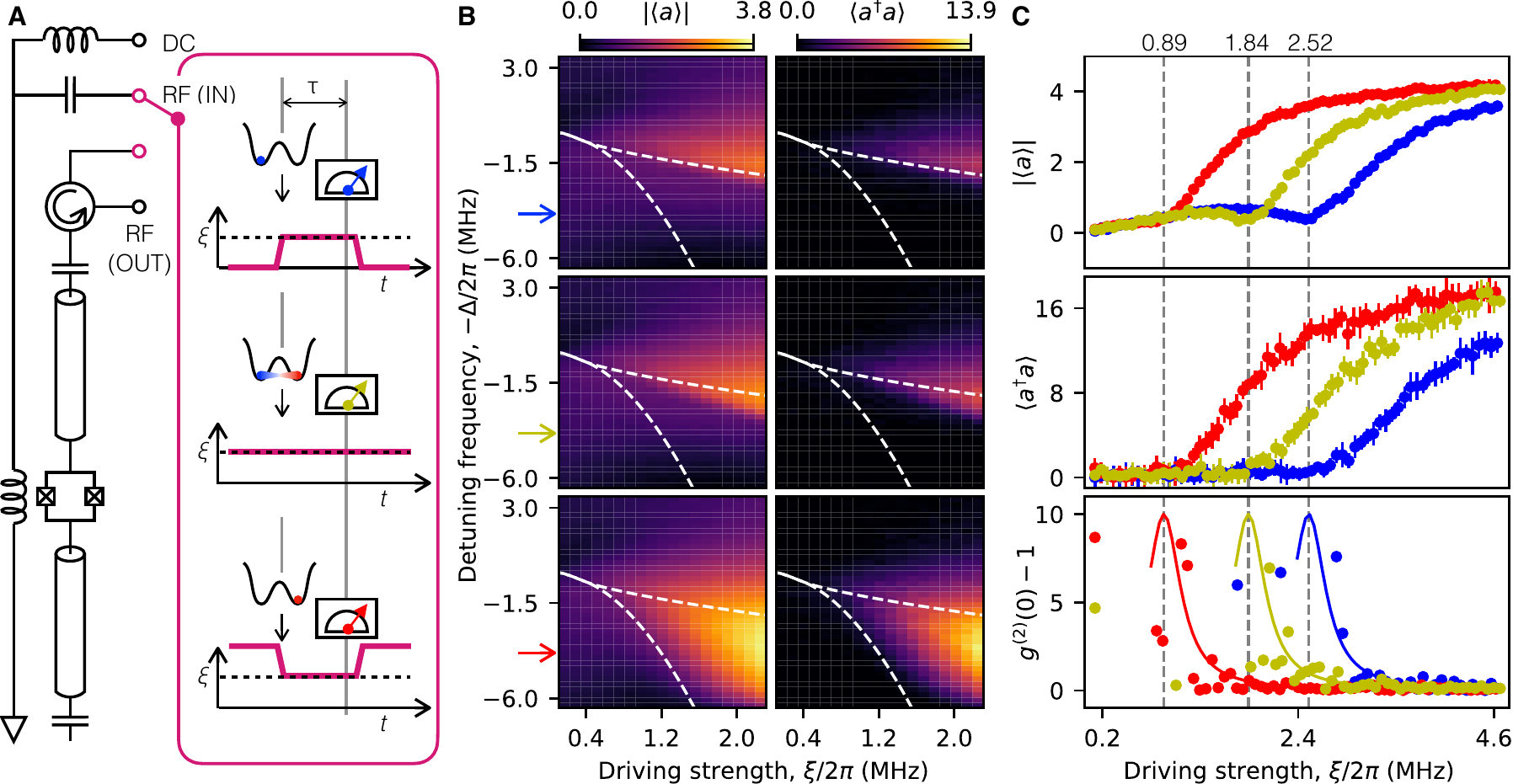}
  \caption{{\bf Hysteresis and its quantum features.} 
{\bf (A)} Schematic of the experimental setup for pulsed heterodyne measurement. The Duffing oscillator is initially prepared in one of the two potential wells (blue and red) or in the SS (yellow). Then, we switch the driving strength to $\xi$, and trigger a short measurement after a waiting time of $\tau$. The DC port is used to control the nonlinearity of the resonator, and we drive the resonator through one of the two RF ports for transmission- or reflection-type measurements. 
{\bf (B)} The absolute mean field, $|\langle a \rangle|$, and photon number, $\langle a^{\dagger}a \rangle$, measured for $\omega_A/2\pi = 7.00\,{\rm GHz}$ show a clear dependance on different initial states in the classical hysteresis regime enclosed by the dashed curves calculated without any fitting parameter \cite{supplement}. A drastic change happens at either of the two boundaries if the system is initially prepared in one well. 
{\bf (C)} At a fixed detuning frequency marked by the arrows in (B), the $|\langle a \rangle|$ vs. $\xi$ curves show a dip around the transition point (vertical dashed lines), while $\langle a^{\dagger}a \rangle$ is a monotonic function of $\xi$. The error bars represent the standard deviation over $8$ independent experiments. The second-order correlation function $g^{(2)}(0)$ is strongly peaked around the transition point, which decays towards unity for large $\xi$. The solid curves are Lorentzian functions serving as guides to the eyes.}
\end{figure}

\newpage
\begin{figure}
  \centering
  \includegraphics[width=5.0in]{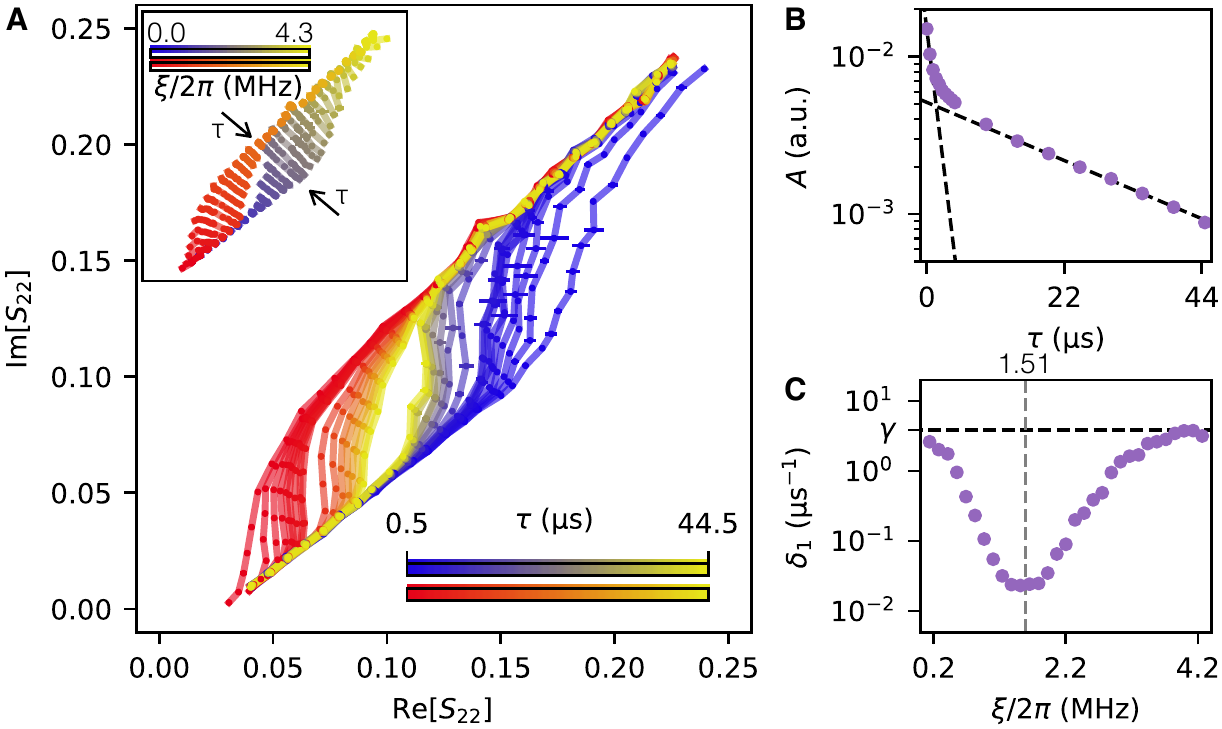}
  \caption{{\bf Two-stage relaxation towards the single unique SS.} 
{\bf (A)} The reflection coefficients, $S_{22}$, corresponding to the two MS branches (blue and red) form a closed loop, which converge to the unique SS solution with $\tau$. The inset shows the convergence of the MSs at each fixed $\xi$.  
{\bf (B)} The loop area, $A$, decays with $\tau$ and shows two distinct decay rates. The dashed lines show the exponential fits, $A\propto \exp (-\eta\tau)$, of the decay rate at small and large $\tau$, with fitted values $\eta_1 = 0.74\,{\rm \mu s^{-1}}$ and $\eta_2 = 0.04\,{\rm \mu s^{-1}}$, respectively.
{\bf (C)} The Liouvillian gap, $\delta_1$, is approximately equal to the total energy dissipation rate, $\gamma$, at a sufficiently small or large $\xi$ (dashed). However, it decreases by more than two orders of magnitude when approaching the critical point, $\xi^{*}/2\pi=1.51\,{\rm MHz}$, and achieves a minimum value of $0.02\,{\rm \mu s^{-1}}$ at $\xi^{*}$. In all panels, the resonant frequency is fixed at $\omega_{\rm A}/2\pi=7.10\,{\rm GHz}$. The error bars represent the standard deviation over $16$ independent experiments, which are smaller than the size of the dots in (B) and (C).}
\end{figure}

\newpage
\begin{figure}
  \centering
  \includegraphics[width=3.5in]{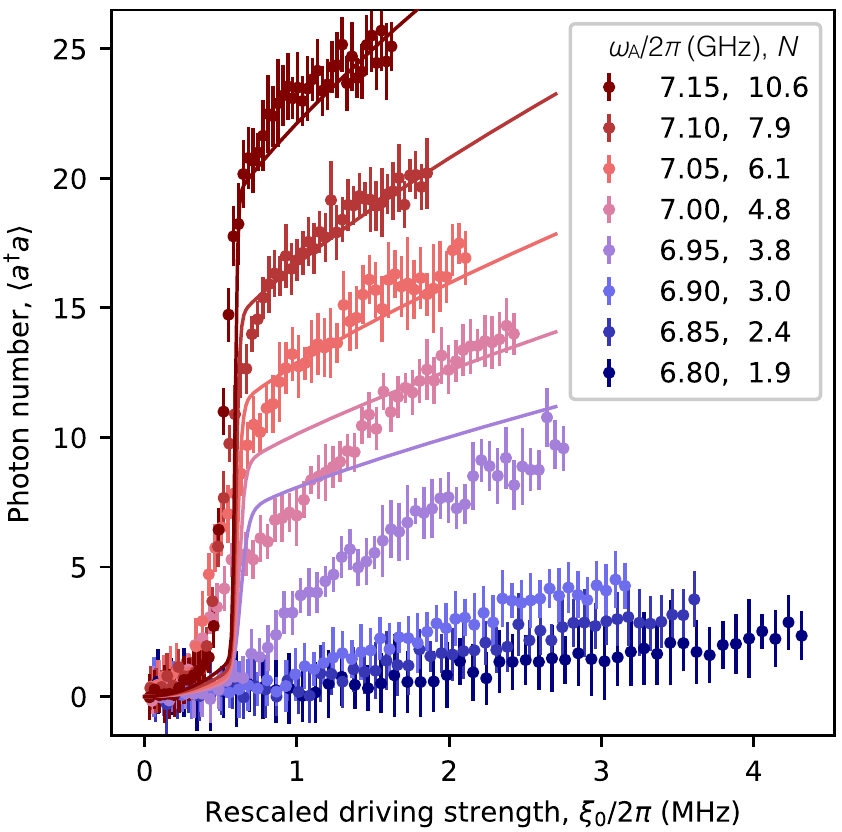}
  \caption{{\bf First-order DPT manifested by increasingly sharp transition step.} 
When approaching the thermodynamic limit ($N\rightarrow \infty$), the observed step of the $\langle a^{\dagger}a \rangle$ vs. $\xi_0$ curve becomes increasingly sharp for a fixed detuning $\Delta=3\gamma$, indicating a first-order DPT. Here, the error bars represent the standard deviation over $8$ independent experiments, and the solid lines are calculated from the quantum theory with no fitting parameter. The deviation between theory and experiment becomes increasingly large at lower resonant frequencies, which we attribute to the increasingly large dephasing rate when tuning the DC flux of the SQUID away from its sweet spot \cite{supplement}.}
\end{figure}

\newpage
\begin{figure}
  \centering
  \includegraphics[width=7.3in]{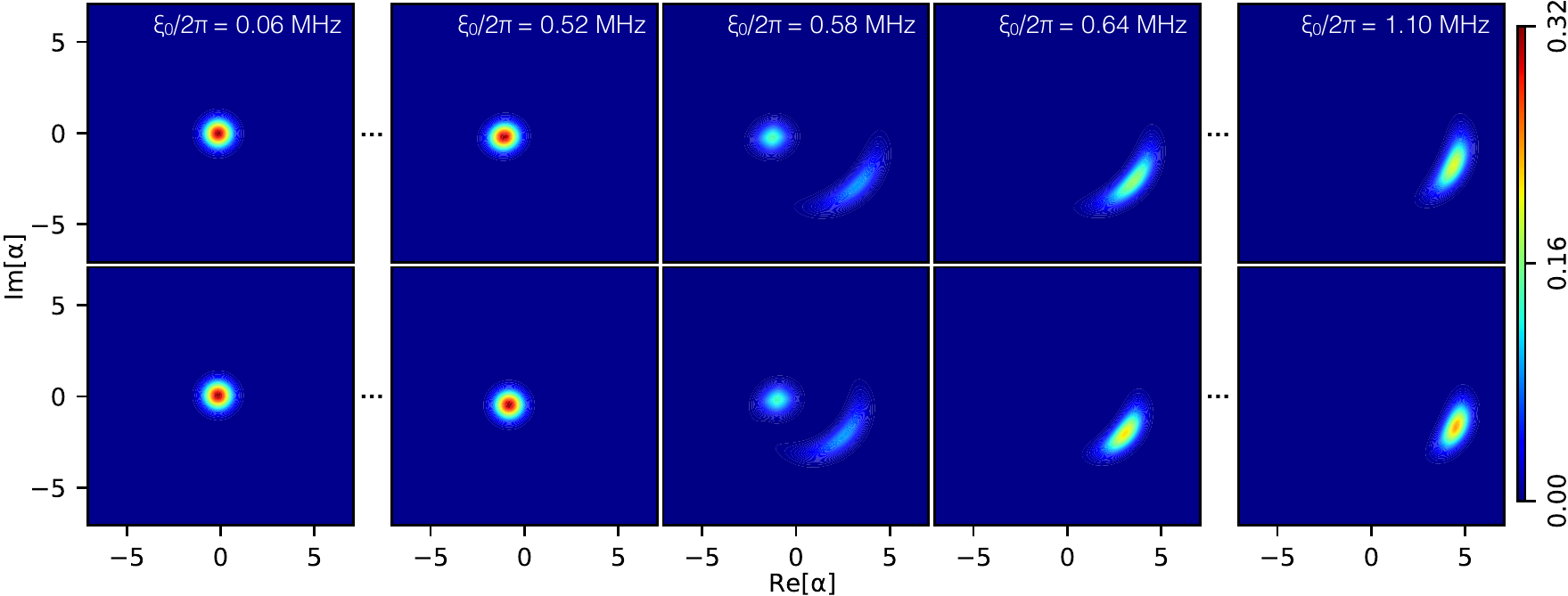}
  \caption{{\bf Wigner function of the SS during DPT.} 
Shown are theory (top) with no fitting parameter and experiment (bottom) for the resonance frequency $\omega_{\rm A}/2\pi=7.15\,{\rm GHz}$. The SS is approximately a coherent (squeezed) state before (after) the phase transition, which separates the two phases of the Duffing oscillator. The transition between the two phases happens within a relatively small range, $0.52\,{\rm MHz} \leq \xi_{0}/2\pi \leq 0.64\,{\rm MHz}$, during which the SS has two distinctive parts in the phase space and is a weighted mixture of the two phases. Ideally, it reaches a equiprobable mixture of the two phases at the exact critical point, which is around $\xi_{0}^{*}/2\pi=0.58\,{\rm MHz}$.}
\end{figure}

\end{document}



\title{Supplemental Materials for \\``Quantum behavior of a superconducting Duffing oscillator at the dissipative phase transition''}
\author{Qi-Ming Chen}
\author{Michael Fischer}
\affiliation{Walther-Mei{\ss}ner-Institut, Bayerische Akademie der Wissenschaften, 85748 Garching, Germany}
\affiliation{Physik-Department, Technische Universit{\"a}t M{\"u}nchen, 85748 Garching, Germany}

\author{Yuki Nojiri}
\author{Michael Renger}
\author{Edwar Xie}
\affiliation{Walther-Mei{\ss}ner-Institut, Bayerische Akademie der Wissenschaften, 85748 Garching, Germany}
\affiliation{Physik-Department, Technische Universit{\"a}t M{\"u}nchen, 85748 Garching, Germany}

\author{Matti Partanen}
\altaffiliation{Present address: IQM, Keilaranta 19, FI-02150 Espoo, Finland}
\affiliation{Walther-Mei{\ss}ner-Institut, Bayerische Akademie der Wissenschaften, 85748 Garching, Germany}

\author{Stefan Pogorzalek}
\altaffiliation{Present address: IQM, Nymphenburger Str. 86, 80335 M{\"u}nchen, Germany}
\affiliation{Walther-Mei{\ss}ner-Institut, Bayerische Akademie der Wissenschaften, 85748 Garching, Germany}
\affiliation{Physik-Department, Technische Universit{\"a}t M{\"u}nchen, 85748 Garching, Germany}

\author{Kirill G. Fedorov}
\affiliation{Walther-Mei{\ss}ner-Institut, Bayerische Akademie der Wissenschaften, 85748 Garching, Germany}
\affiliation{Physik-Department, Technische Universit{\"a}t M{\"u}nchen, 85748 Garching, Germany}

\author{Achim Marx}
\affiliation{Walther-Mei{\ss}ner-Institut, Bayerische Akademie der Wissenschaften, 85748 Garching, Germany}

\author{Frank Deppe}
\altaffiliation{Present address: IQM, Nymphenburger Str. 86, 80335 M{\"u}nchen, Germany}
\affiliation{Walther-Mei{\ss}ner-Institut, Bayerische Akademie der Wissenschaften, 85748 Garching, Germany}
\affiliation{Physik-Department, Technische Universit{\"a}t M{\"u}nchen, 85748 Garching, Germany}
\affiliation{Munich Center for Quantum Science and Technology (MCQST), 80799 Munich, Germany}

\author{Rudolf Gross}
\affiliation{Walther-Mei{\ss}ner-Institut, Bayerische Akademie der Wissenschaften, 85748 Garching, Germany}
\affiliation{Physik-Department, Technische Universit{\"a}t M{\"u}nchen, 85748 Garching, Germany}
\affiliation{Munich Center for Quantum Science and Technology (MCQST), 80799 Munich, Germany}

\date{\today}
\maketitle
\tableofcontents

\phantom{\cite{Nayfeh1995, Landau1976, Siddiqi2004, Siddiqi2005, Gibbs1976, Rempe1991, Yurke1984, Lin2014, Fajans2001, Murch2010, Drummond1980, Naaman2008, Mavrogordatos2017, Brookes2021, Rodriguez2017, Fink2017, Macieszczak2016, Casteels2017, Minganti2018, supplement, Albert2014, Leib2012, Fischer2021, Jung1990, Casteels2016, Bajer2004, Vogel1989, Kheruntsyan1999, Raftery2014, Fitzpatrick2017, Ma2019, Carusotto2013}}

\thispagestyle{empty}

\clearpage
\section{Model and theory}
\setcounter{page}{1}
\subsection{Description of the system}\label{sec:sample}
\begin{figure}[hbt]
  \centering
  \includegraphics[width=0.8\columnwidth]{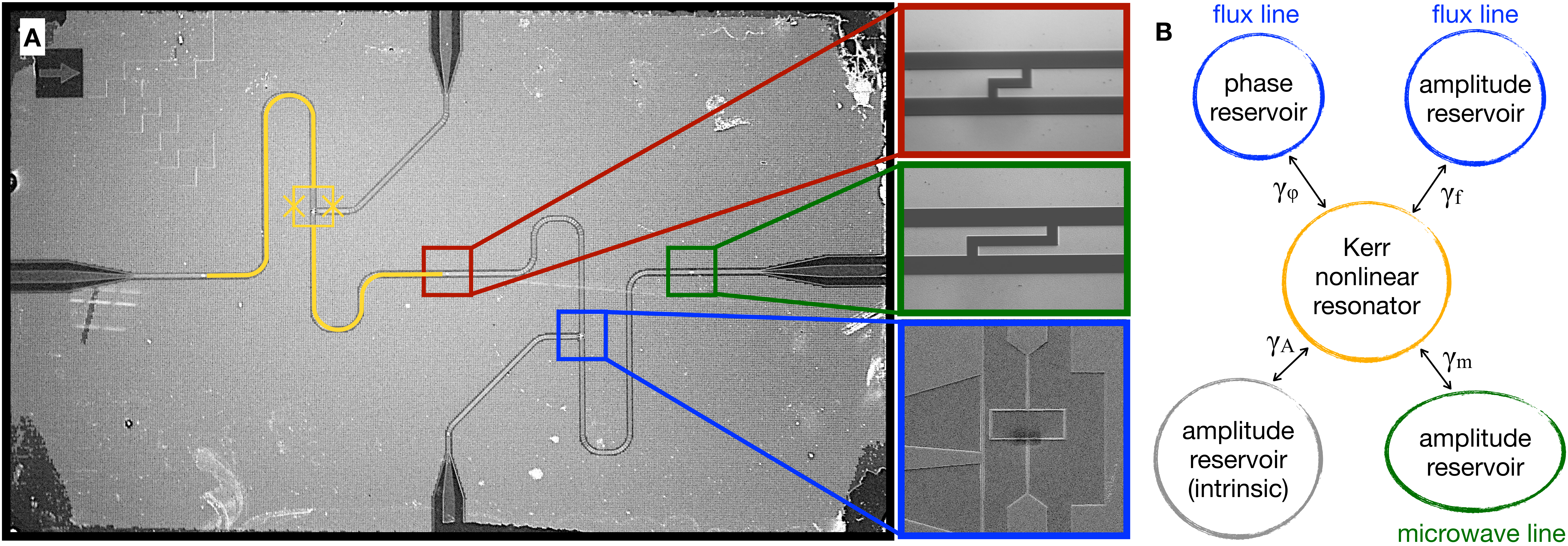}
  \caption{{\bf Optical photograph of the sample chip.} The sample consists of two transmission line resonators with a DC-SQUID embedded in the middle (yellow). The two resonators are coupled by a finger capacitor, as shown in the top red box. They are also coupled to two external feedlines, as shown in the middle green box. The resonant frequency and the nonlinearity of each individual resonator is controlled by the T-shaped flux control lines, as shown in the bottom blue box.}
  \label{fig:sample}
\end{figure}

An optical photograph of a reference sample is shown in Fig.\,\ref{fig:sample}A, which has the same design as that used in the experiment. The sample is fabricated on a $525\,{\rm \mu m}$-thick silicon chip with an area of $10\times 6\,{\rm mm^2}$ using double-angle shadow evaporation and lift-off procedures. The superconductor layer is made of aluminum with a thickness of $140\,{\rm nm}$. The major part of the sample consists of two $7.2\,{\rm mm}$-long and $13.2\,{\rm \mu m}$-wide transmission line resonators with two DC-SDUIDs embedded in the middle, respectively. The areas of the two SQUIDs are designed to be $10.5\times 24.5\,{\rm \mu m^2}$ and the two junctions in the SQUID loop differ in size to achieve a SQUID asymmetry of approximately $0.13$. In addition, two T-shaped on-chip antennae are placed in proximity to the two SQUIDs, respectively, to control the magnetic flux threading the SQUID loops. With this sample design we achieve two nonlinear resonators with tunable frequency and nonlinearity \cite{Leib2012, Fischer2021}. The two resonators are coupled by a $20\,{\rm \mu m}$-long finger capacitor. Furthermore, they are coupled to the outside fields, respectively, by two $40\,{\rm \mu m}$-long finger capacitors at the two ends, and also to the microwave fields in the flux control lines through the two antennae.

In our experiment, we focus on a single resonator, which is labelled as ``Resonator-$2$" in the rest of the context. The other resonator is detuned by at least $100\,{\rm MHz}$ throughout our experiment, which is much larger than the coupling strength between the two resonators ($\sim 5\,{\rm MHz}$) and thus can be fairly neglected. Fig.\,\ref{fig:sample}B shows the schematic of the whole system, of which the Hamiltonian is described as
\begin{align}
	H/\hbar &= \omega_{\rm A} a^{\dagger}a + Ua^{\dagger}a^{\dagger}aa 
	+ \sum_{k=-\infty}^{+\infty} \omega_{k} b_{{\rm A},k}^{\dagger}b_{{\rm A},k}
	+ i\kappa_{\rm A} \left( b_{{\rm A},k}^{\dagger}a - b_{{\rm A},k}a^{\dagger} \right) \nonumber \\
	&+ \sum_{k=-\infty}^{+\infty} \omega_{k} b_{{\rm m},k}^{\dagger}b_{{\rm m},k} 
	+ i\kappa_{\rm m} \left( b_{{\rm m},k}^{\dagger}a - b_{{\rm m},k}a^{\dagger} \right) \nonumber \\
	&+ \sum_{k=-\infty}^{+\infty} \omega_{k} b_{{\rm f},k}^{\dagger}b_{{\rm f},k} 
	+ i\kappa_{\rm f} \left( b_{{\rm f},k}^{\dagger}a - b_{{\rm f},k}a^{\dagger} \right) 
	+ i\kappa_{\rm \varphi} \left( b_{{\rm f},k}^{\dagger} - b_{{\rm f},k} \right)a^{\dagger}a. 
\end{align}
Here, $a$, and $b_{{\rm A/m/f},k}$ are the field operators of the resonator and the intrinsic, microwave-line, and flux-line reservoirs, respectively. The parameter $\kappa_{\rm A/m/f/\phi}$ describes the coupling strength between the system, i.e., the resonator, and the corresponding reservoirs. Following the standard derivation of the input-output formalism \cite{Chen2021,Chen2021b} and restricting our discussion to a narrow bandwidth around the driving frequency, $\omega_{\rm d}$, we obtain the following Heisenberg-Langevin equation for the resonator degree of freedom
\begin{align}
	\dot{a}(t) &= -i\omega_{\rm A} a(t) -i2Ua^{\dagger}(t)a^2(t) 
	- \frac{\gamma_{\rm A}+\gamma_{\rm m}+\gamma_{\rm f}}{2}a(t) 
	- \sqrt{\gamma_{\rm m}}b_{\rm in}^{({\rm m})}(t)
	- \sqrt{\gamma_{\rm f}}b_{\rm in}^{({\rm f})}(t)
	- \sqrt{\gamma_{\rm \varphi}}a(t)\left(b_{\rm in}^{({\rm f})}(t)-b_{\rm in}^{({\rm f})\dagger}(t)\right),
	\label{eq:langevin}
\end{align}
where $\sqrt{\gamma_{\rm A/m/f/\varphi}}=\sqrt{2\pi}\kappa_{\rm A/m/f/\varphi}$. Here, we have assumed the intrinsic amplitude reservoir to be in the vacuum state, and omitted the two-photon loss term by rotating wave approximation (RWA). The input fields are defined as $b_{\rm in}^{({\rm m/ f})}(t)=\int_{-\infty}^{+\infty} d\omega e^{-i\omega t}b_{\omega}^{({\rm m/ f})}(0)/\sqrt{2\pi}$, while the output field in the microwave line is $b_{\rm out}^{({\rm m})}(t) = b_{\rm in}^{({\rm m})}(t) + \sqrt{\gamma_{m}}a(t)$.

\subsection{Quantum theory of the Duffing oscillator}\label{sec:duffing_quantum}
Because our experiments are performed in the regime where the dephasing rate, $\gamma_{\phi}$, is smaller than the total energy dissipation rate, $\gamma=\gamma_{\rm A}+\gamma_{\rm m} + \gamma_{\rm f}$, we temporarily omit the dephasing effect in the following discussions. The experimentally determined values of the sample parameters are reported in Section\,\ref{sec:open}. We will discuss the dephasing effect and also the possible two-photon processes in Section\,\ref{sec:dephasing} for achieving a better understanding between the experimental data and the simulation results. In the rotating frame at the driving frequency, $\omega_{\rm d}$, we obtain the simplified Heisenberg-Langevin equation as
\begin{align}
	\dot{a}(t) &= -i\Delta a(t) -i2Ua^{\dagger}(t)a^2(t) 
	- \frac{\gamma}{2}a(t) 
	- \xi. \label{eq:duffing_langevin}
\end{align}
Here, $\Delta = \omega_{\rm A}-\omega_{\rm d}$ is the frequency detuning between the resonator and the drive. Besides, we have assumed the input field to be coherent such that we can use a complex number to describe the driving strength, $\xi=-i\sqrt{\gamma_{\rm m/f}}\langle b_{\rm in}^{(\rm m/f)} \rangle$ \cite{Haroche2006}. In most measurements reported in the main text, we drive the sample through the flux line while measuring through the microwave line, in order to avoid the reflecting driving field in the output path. 

The above equation describes a quantum-mechanical Duffing oscillator. It has been proven that all orders of signal moments for the steady state (SS) can be calculated in an analytical way \cite{Drummond1980}
\begin{align}
	\langle a^{\dagger j}a^k \rangle 
	&= d^{*j}d^{k}
	\frac{\Gamma(c)\Gamma(c^*){}_0\mathcal{F}_2\left(k+c,j+c^*, 2\left|d\right|^2 \right)}
	{\Gamma(k+c)\Gamma(j+c^*){}_0\mathcal{F}_2\left(c,c^*, 2\left|d\right|^2 \right)}. \label{eq:positive_p}
\end{align}
Here, we have used the abbreviations $c=\left(\Delta - i\gamma/2 \right)/U$ and $d=-\xi/U$. Moreover, ${}_0\mathcal{F}_2\left(x,y,z\right)=\sum_{n=0}^{\infty}\Gamma(x)\Gamma(y)z^n/\left[\Gamma(x+n)\Gamma(y+n)n!\right]$ is a generalized hypergeometric function, where $\Gamma(\cdot)$ is the gamma special function. This formula indicates that the signal moments of the steady states (SSs) are single valued in the entire parameter space, such that a quantum-mechanical Duffing oscillator ``does not exhibit bistability or hysteresis" \cite{Drummond1980}. Theoretical calculations also indicate the following Wigner quasi-distribution of the unique SS \cite{Vogel1989, Kheruntsyan1999}
\begin{align}
	W\left(\alpha,\alpha^*\right) = \mathcal{N}e^{-2\left|\alpha\right|^2}\left|{}_{0}\mathcal{F}_{1}\left(c, 2d\alpha^{*} \right)\right|^2,
	\label{eq:exact_solution}
\end{align}
where $\mathcal{N}$ is a normalization factor and ${}_{0}\mathcal{F}_{1}(x,z)=\sum_{n=0}^{\infty}\Gamma(x)z^n/\left[\Gamma(x+n)n!\right]$ is a hypergeometric function.

\subsection{Classical theory of the Duffing oscillator}\label{sec:duffing_classical}
If we take the mean value of the Heisenberg-Langevin equation in Eq.\,\eqref{eq:duffing_langevin}, and neglect the photon correlations in the third-order term, i.e., $\langle a^{\dagger}(t)a^2(t) \rangle \rightarrow \alpha^{*}(t)\alpha^2(t)$ where $\alpha(t)=\langle a \rangle$, we obtain the classical equation of motion for the Duffing oscillator \cite{Casteels2017}
\begin{align}
	\dot{\alpha}(t) &= -i\Delta \alpha(t) -i2U\alpha^{*}(t)\alpha^2(t) - \frac{\gamma}{2}\alpha(t) 
	- \xi(t). \label{eq:duffing_classical}
\end{align}
One can prove that this equation is equivalent to the celebrated Duffing equation under RWA \cite{Dykman2007, Serban2010, Guo2011}. The SS solution of the classical system can be obtained by solving the following equation \cite{Landau1976, Drummond1980}
\begin{align}
	4U^2 \left|\alpha\right|^6 + 4\Delta U\left|\alpha\right|^4 + \left[\left(\frac{\gamma}{2}\right)^2 + \Delta^2 \right]\left|\alpha\right|^2 - \left|\xi\right|^2 = 0. \label{eq:mean_field}
\end{align}
Depending on the specific parameter settings, either one, two, or three solutions of $|\alpha|^2$ are allowed in the so-called hysteresis regime. This observation is in stark contrast to the quantum-mechanical analysis, where a single unique SS solution is predicted throughout the entire parameter space. The stability of the system can be verified by checking whether $\partial \left|\xi\right|^{2}/\partial |\alpha|^2 > 0$ \cite{Landau1976,Drummond1980}. At the boundaries of the hysteresis regime, we have
\begin{align}
	\left|\alpha\right|^2 = \frac{-2\Delta \pm \sqrt{\Delta^2-3\left(\gamma/2\right)^2}}{6U},
\end{align}
which, in combination with Eq.\,\eqref{eq:mean_field}, can be used to draw the two boundaries of the hysteresis regime in the $\xi$-$\Delta$ space, as shown in Fig.\,1B of the main text. It indicates that the hysteresis and bistability exist only in the regime $\Delta^2 > 3\left(\gamma/2\right)^2$, where the system has a double-well potential. Outside this parameter regime, the potential has only a single minimum and there exists only a single unique SS solution. Depending on whether the driving strength is smaller or larger than either of the two boundaries, the single well is approximately localized at one of the two minima of the double-well potential. This observation leads to the method of initial state preparation, as will be discussed in detail in Section\,\ref{sec:methods}. 

\section{System characterization}\label{sec:setup}
\subsection{Experimental setup}\label{sec:cryo_setup}
\begin{figure}[hbt]
  \centering
  \includegraphics[width=\columnwidth]{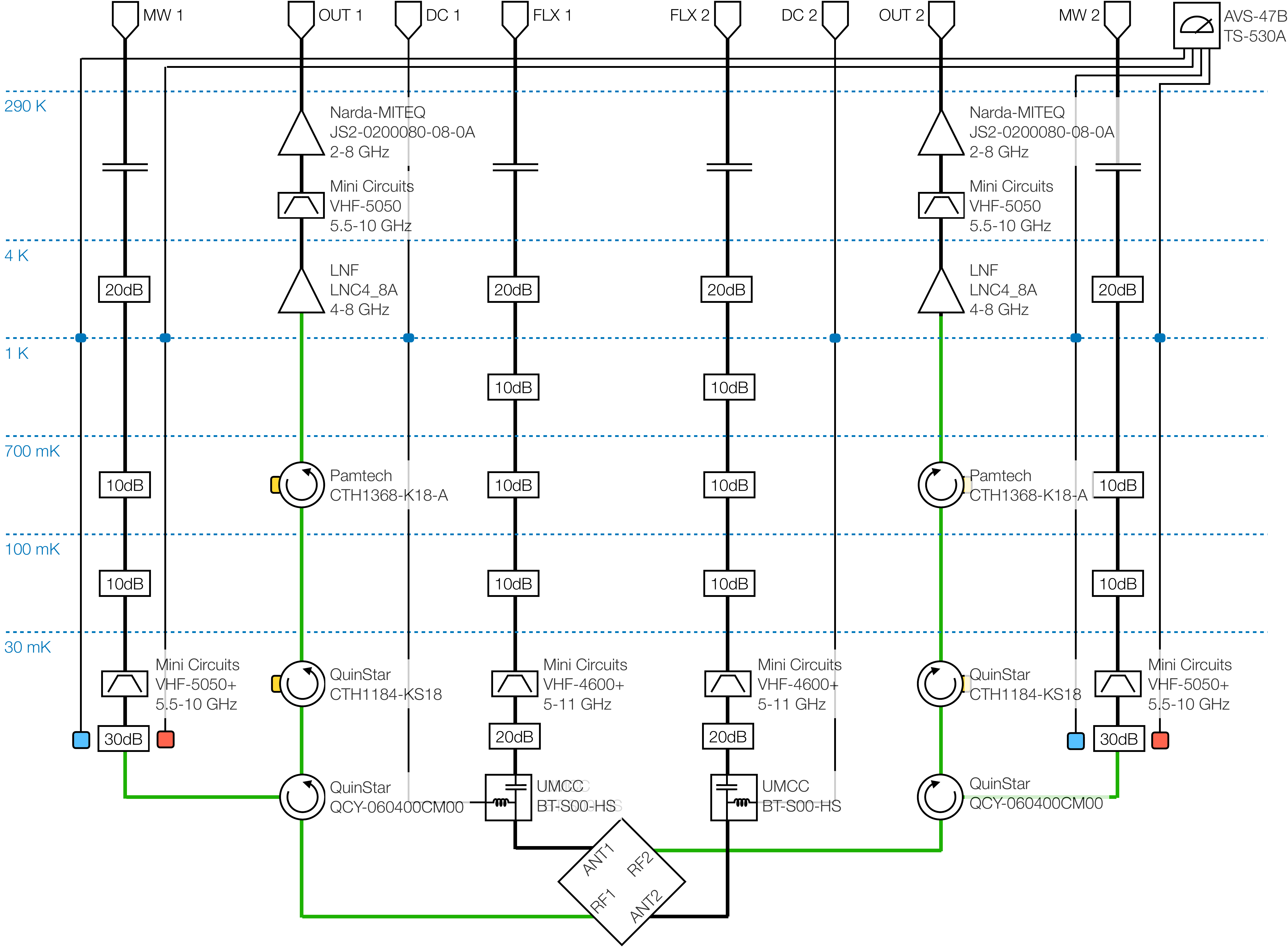}
  \caption{{\bf Schematic of the experimental setup.} The sample is placed at the mixing chamber stage of a homemade wet-type dilution refrigerator, to which four microwave coaxial cables are connected. We also anchor two pairs of homemade cryogenic thermometers (blue box) and heaters (red box) to the two $30\,{\rm dB}$ attenuators at the base temperature to realize active control of the local temperature. Here, the blue dashed lines indicate the temperature stages of the cryostat, and the blue dots indicate a heat exchanger for cooling the DC wires. The green thick lines indicate NbTi cables.}
  \label{fig:cryo_setup}
\end{figure}

The schematic of the experimental setup that is placed inside, or integrated on top of the dilution refrigerator is shown in Fig.\,\ref{fig:cryo_setup}. From the top to the bottom, the temperature decreases from approximately $290\,{\rm K}$, which is stabilized by using the Peltier cooler (Laird Hi-Pot tested 750VOC) and the temperature stabilizer (Telemeter TR12-PI-2Q2), to a minimum value of $30\,{\rm mK}$ at the sample stage. The input and output microwave lines, labelled as MW\,1/2, FLX\,1/2, and OUT\,1/2, are coupled to the two nonlinear resonators through the on-chip finger capacitors and the T-shaped antennae, as shown in Fig.\,\ref{fig:sample}. Here, the microwave fields in MW\,1/2 and OUT\,1/2 are separated by using the cryogenic circulators (QuinStar QCY-060400CM00). We add also a $5.5$-$10\,{\rm GHz}$ high-pass filter in each of the input paths to isolate the sample from higher-frequency harmonics of the driving fields. In each of the the output paths, we add two circulators (QuinStar CTH1184-KS18, Pamtech CTH1368-K18-A) at $30\,{\rm mK}$ and $700\,{\rm mK}$, respectively, to isolate the sample from the high-temperature thermal radiations and the possible back propagating fields coming from the HEMT amplifiers (LNC4\_8A). At the top of the cryostat, we place a $5.5$-$10\,{\rm GHz}$ high-pass filter and amplify the cryogenic signal by a low-noise room-temperature amplifier (MITEQ JS2-0200080-08-0A) in each of the output line. These amplifiers are tightly integrated with the Peltier cooler, such that they operate at a stable  temperature of around $17\,{\rm {}^{\circ}C}$. We place all of the described microwave components in an electromagnetically shielded room, while the temperature of the entire laboratory is stabilized around $27\,{\rm {}^{\circ}C}$ by using the air conditioner.

For the DC part, we combine the output of the DC current sources (ADCMT 6241A) with the microwave fields in FLX\,1/2 by using the bias-tee (UMCC BT-S00-HS), which are further connected to the T-shaped antennae on chip. In addition, two pairs of homemade cryogenic thermometers and heaters are clamped tightly to the two $30\,{\rm dB}$ attenuators at base temperature. They are connected to a AC resistance bridge (Picowatt AVS-47B) and the corresponding PID temperature controller (Picowatt TS-530A), in order to control the local temperatures of the two attenuators and generate blackbody radiation. This configuration is used to characterize the amplification gain and the noise temperature of the output paths, as discussed in Section\,\ref{sec:output}.

\subsection{Control and readout modules}\label{sec:rt_setup}
\begin{figure}[hbt]
  \centering
  \includegraphics[width=0.95\columnwidth]{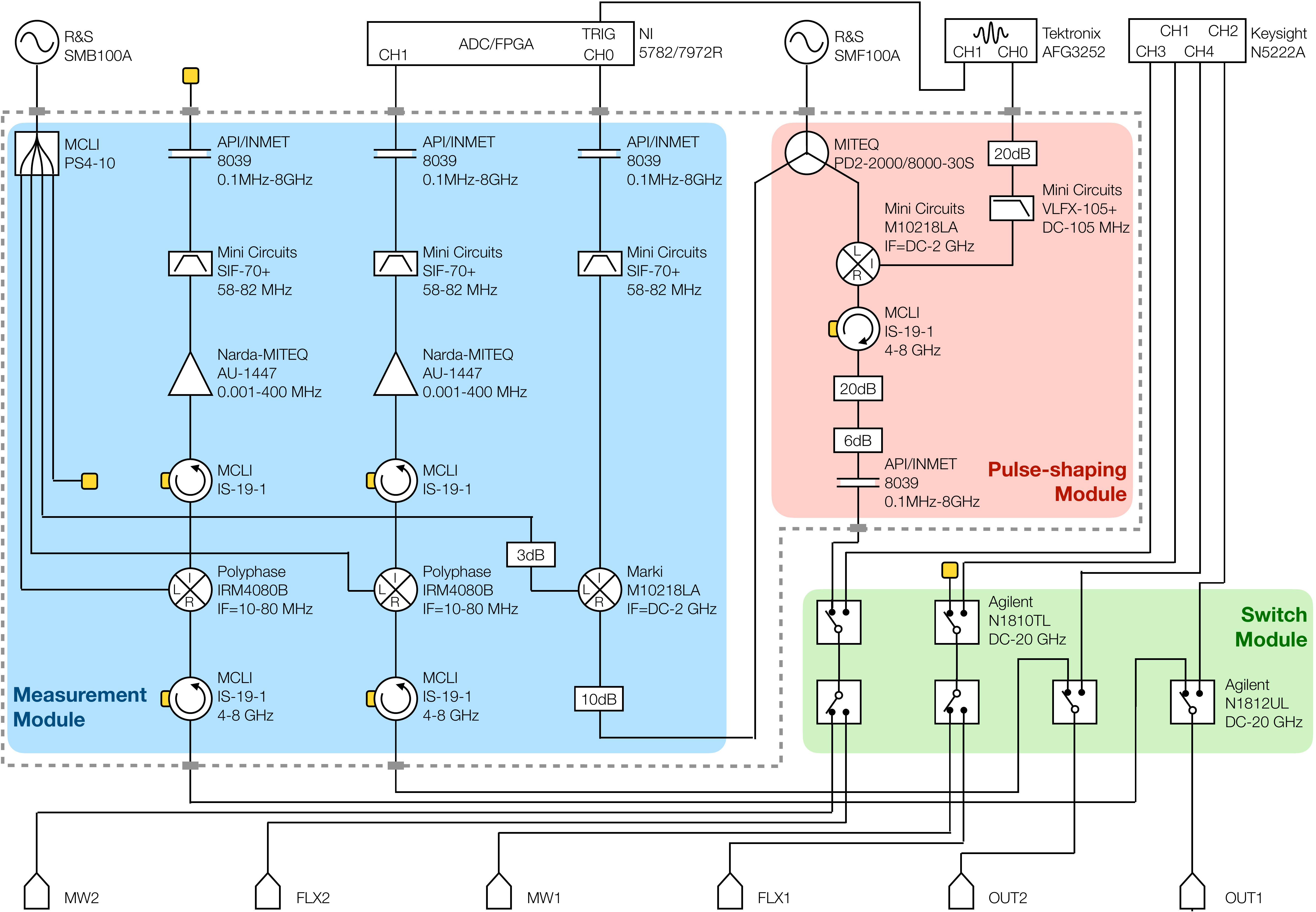}
  \caption{{\bf Schematic of the control and readout modules.} The setup consists of three modules: The measurement module (blue), the pulse-shaping module (red), and the switch module (green). The components enclosed by the dashed lines are sealed in a $48\times 24\times 12\,{\rm cm^2}$ homemade copper box for electromagnetic shielding, grounding, and passive cooling. The switch module is placed outside the box to avoid potential stray magnetic field that may influence the other microwave devices.}
  \label{fig:rt_setup}
\end{figure}

The schematic of the room-temperature setup for control and measurement is shown in Fig.\,\ref{fig:rt_setup}, which consists of three modules. The pulse-shaping module is designed to control the initial state of the nonlinear resonator and also to drive the system. We use a microwave signal generator (R\&S SMF100A) to generate the radio frequency (RF) carrier wave. The field envelope is modulated by a double balanced mixer (Marki M1-0218LA) with its local oscillator (LO) port connected to the carrier wave and the intermediate frequency (IF) port to the first channel of an AFG(arbitrary function generator, Tektronix AFG3252). The RF port of the mixer is connected to the switch module for further signal routing. The second channel of the AFG is synchronized with the first one, which is used to trigger the measurement process of the ADC (analogue-to-digital converter, NI FlexRIO 5782). In addition, we place several attenuators, circulators, and filters in the configuration for the compatibility of different microwave devices.

The measurement module is designed to down-convert the RF signal to an IF frequency of $f_{\rm IF}=62.5\,{\rm MHz}$ for pulsed heterodyne measurements. This choice of frequency avoids the possible beating between the signal and the higher order harmonics of the $10\,{\rm MHz}$ Rb frequency standard (SRS FS725), which synchronizes all the instruments in the lab. We use image rejection mixers (Polyphase IRM4080B) in the first two lines, OUT\,1/2, to achieve a better signal-to-noise ratio (SNR), while a double balanced mixer (Marki M1-0218LA) is used in the third reference line for its relatively low price. However, we use the same LO field, which is generated by the microwave signal generator (R\&S SMB100A), to drive all the three mixers for reaching a phase alignment. We also amplify the two channels, OUT\,1/2, by low-noise room-temperature amplifiers (MITEQ AU1447R), and place several attenuators, filters, isolators, power dividers to improve the SNR. We note that the isolator (MCLI IS-19-1) is designed for the $4$-$8\,{\rm GHz}$ range, while it still works in the megahertz regime for our needs of preventing the possible back propagating fields from the IF amplifier.

Besides, we use several microwave coaxial switches (Agilent N1810TL, N1812UL) in the switch module to control the connectivity of different signal paths for different experimental purposes. The switches are controlled by a commercial controller (Agilent L4445A) with a homemade remote-control panel. For typical characterization experiments, where only the scattering coefficients are measured, we connect the two input ports, MW\,1/2, and the two output ports, OUT\,1/2, to the four channels of the VNA (vector network analyzer, Keysight PNA N5222A). However, for quadrature measurements we connect OUT\,2 and the reference driving field to the two channels of the ADC, which has a sampling frequency of $f_{\rm S}=250\,{\rm MHz}$. The driving field is also connected to MW\,2 for reflection-type measurements, or FLX\,2 for transmission-type measurements. 

\subsection{Closed-system parameters}\label{sec:closed}
\begin{table}[hbt]
  \centering
  \begin{tabular}{|l|l|}
    \hline
    CPW resonator & \\
    \hline
    length $L$ & $7.395\times 10^{-3}\,{\rm m}$ \\
    \hline
    inductance per meter $l$ & $4.598\times 10^{-7}\,{\rm H}$ \\
    \hline
    capacitance per meter $c$ & $1.697\times 10^{-10}\,{\rm F}$ \\
    \hline\hline
    SQUID \#1, \#2 & \\
    \hline
    critical current $I_{J1}$, $I_{J2}$ & $1.566\times 10^{-6}\,{\rm A}$, $1.416\times 10^{-6}\,{\rm A}$ \\
    \hline
    shunting capacitance $C_{J1}$, $C_{J2}$ & $9.394\times 10^{-16}\,{\rm F}$, $1.168\times 10^{-15}\,{\rm F}$ \\
    \hline
    asymmetry $d_{J1}$, $d_{J2}$ & $2.136\times 10^{-1}$, $1.937\times 10^{-1}$ \\
    \hline\hline
    Antenna \#1, \#2 & \\
    \hline
    flux offset $\phi_{1, \rm off}$, $\phi_{2, \rm off}$ & $-3.902\times 10^{-1}\,{\phi_0}$, $-1.149\times 10^{-1}\,{\phi_0}$ \\
    \hline
    flux change per current $d\phi_1/dI_1$, $d\phi_2/dI_1$ & $6.088\times 10^{-4}\,{\rm \phi_0/A}$, $9.927\times 10^{-4}\,{\rm \phi_0/A}$ \\
    \hline
    flux change per current $d\phi_1/dI_2$, $d\phi_2/dI_2$ & $-3.715\times 10^{-5}\,{\rm \phi_0/A}$, $-5.054\times 10^{-4}\,{\rm \phi_0/A}$ \\
    \hline
  \end{tabular}
  \caption{{\bf Experimentally determined closed-system parameters of the system.}}
  \label{tab:autotune}
\end{table}

To determined the closed-system parameters, such as the resonant frequency, $\omega_{\rm A}$, and nonlinearity, $U$, we slowly sweep the current in either of the two antennae and measure the scattering coefficients. Here, we assume a linear relation between the flux and the applied currents \cite{Fischer2021}
\begin{align}
	\left(\begin{matrix}
  \phi_{1}\\
  \phi_{2} 
\end{matrix}\right) = \left(\begin{matrix}
  A_{11} & A_{12} \\
  A_{21} & A_{22} 
\end{matrix} \right)\left(\begin{matrix}
  I_{1}\\
  I_{2} 
\end{matrix}\right) + \left(\begin{matrix}
  \phi_{1, {\rm off}}\\
  \phi_{2, {\rm off}} 
\end{matrix}\right),
\label{eq:cross_talk}
\end{align}
where $A$ is the crosstalk matrix, $\phi_{1, {\rm off}}$ and $\phi_{2, {\rm off}}$ are the offset flux threading into the two SQUID loops, and $\phi_{0}=\hbar/2e$ is reduced flux quantum. We sweep respectively the two antenna currents from $-600\,{\rm mA}$ to $600\,{\rm mA}$ with $80$ intermediate steps, and measure the scattering responses of the system by using the VNA. The sweeping speed is set to $1\,{\rm mA/s}$, and the IF bandwidth of the VNA is set to $1\,{\rm kHz}$. We average each data point over $100$ times with a point-average mode. In principle, the output power of VNA should be set as low as possible in order to minimize the nonlinear effects \cite{Watanabe2009}. However, considering the practical compromise between the SNR and the measurement time, we set the output power to $0\,{\rm dBm}$ with an additional $30\,{\rm dB}$ attenuation at room temperature. The cables inside the cryostat contribute to an $\sim 100\,{\rm dB}$ attenuation, which will be characterized in section\,\ref{sec:input}. This configuration corresponds to an approximately $10\,{\rm h}$ measurement time for each characterization. 

After getting the measurement results, we use an optimization method to find the best estimation of the $9$ closed-system parameters of the sample, as well as the $6$ parameters defined in Eq.\,\eqref{eq:cross_talk} for controlling the external flux. The automated fitting procedure not only provides the possibility to find the optimal estimation of the sample parameters, not matter local or global, in the huge parameter space, but also avoids possible biases in manual characterization and keeps a relatively objective criteria among different experiments. The characterization results are summarized in Table\,\ref{tab:autotune}, which has been reported in Ref.\,\onlinecite{Fischer2021}. The slight difference of several parameters in Table\,\ref{tab:autotune} and Ref.\,\onlinecite{Fischer2021} is attributed to the drift of sample parameters in different cool down.

\subsection{Open-system parameters}\label{sec:open}
Using the automated sample tuning procedure, we tune the second resonator to different frequencies and measure the reflection coefficient for characterizing the open-system parameters, such as the total energy dissipation rate, $\gamma$. The measurement data is processed by the recipe described in Ref.\,\onlinecite{Chen2021}, where the experimental imperfections, such as acquisition noise and circuit asymmetries, are corrected automatically. We note that the reflection coefficient of the effective one-resonator system is slightly different from a typical necklace-type $\lambda/2$ resonator, because we consider only the input and output fields at one single end of the resonator. In other words, we attribute all the photon-loss mechanisms, which include the resonator intrinsic loss as well as the photon loss through the flux line and the resonator-resonator coupling capacitor, into the internal quality factor, $Q_{\rm i}$. Considering also the practical distortions of the spectrum, we write \cite{Fischer2021}
\begin{align}
	S_{22}\left(\omega\right) \approx A e^{-j\left(\tau\omega + \varphi\right)} \left( 1 - \frac{e^{j\phi}2Q_{\rm l}/\left| Q_{\rm c}\right|}{1+2jQ_{\rm l}\left( \omega/\omega_{\rm r} -1 \right)} \right),
\end{align}
Here, $1/Q_{\rm l}=1/Q_{\rm i}+1/Q_{\rm c}$, and we have defined the reflection coefficient of the second resonator as $S_{22}$. To minimize the influence of the resonator nonlinearity and obtain a faithful characterization, we keep $P_{\rm in}$ sufficiently small. We set the power at the VNA output to $-30\,{\rm dBm}$ and add $30$ -- $50\,{\rm dB}$ room-temperature attenuations depending on the SNR at different frequencies. In this way, the spectrum is kept approximately Lorentzian such that the contribution of nonlinearity to the scattering coefficient can be fairly neglected \cite{Watanabe2009}. 

\begin{figure}[hbt]
  \centering
  \includegraphics[width=\columnwidth]{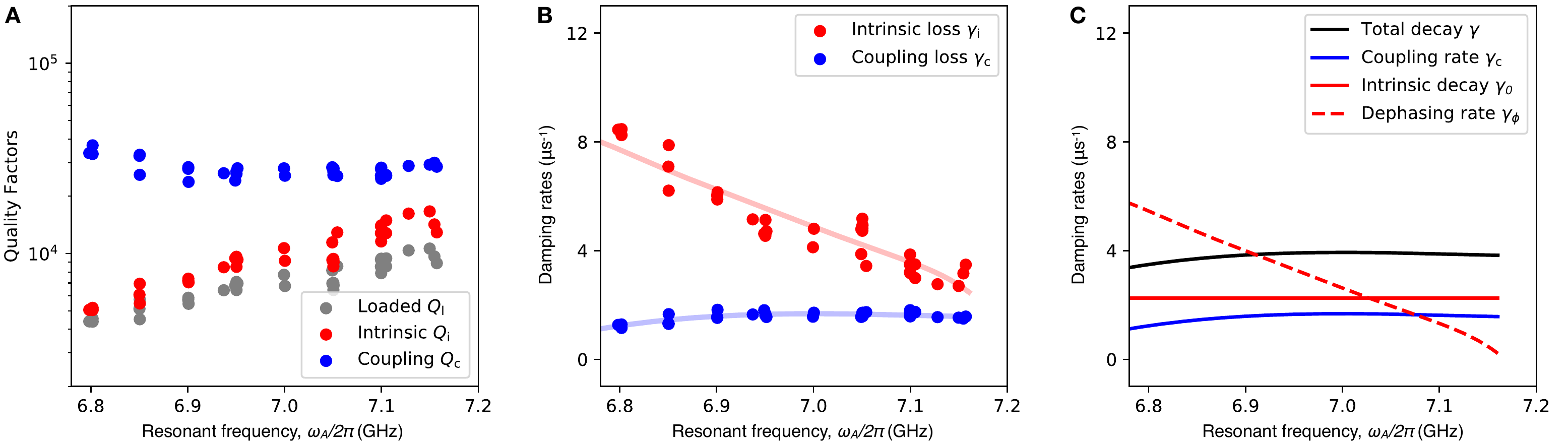}
  \caption{{\bf Experimentally determined open-system parameters of the two resonators.} {\bf (A)} We tune the second resonator to different resonant frequencies and characterize the internal, external, and loaded Q factors. {\bf (B)} The result is transformed into the internal and external loss rates, $\gamma_{i}$ and $\gamma_{\rm m}$, respectively. {\bf (C)} We separate the frequency-dependent and independent parts of $\gamma_{\rm i}$ into the energy decay rate, $\gamma_{0}$, and dephasing rate, $\gamma_{\phi}$. The former can be further split into $\gamma_{\rm A}$ and $\gamma_{\rm f}$.}
  \label{fig:damping_rates}
\end{figure}

Figure\,\ref{fig:damping_rates} summarizes the characterization results of the second resonator in the range between $6.80$ and $7.20\,{\rm GHz}$. Although the external Q factor is approximately a constant for different $\omega_{\rm A}$, the internal Q factor decreases when decreasing the resonant frequency. We note that similar observations are also reported in the literature \cite{PalaciosLaloy2008, Sandberg2008}. We attribute the change of $Q_{\rm i}$ to the possible effect of dephasing, which originates from the jitter of the resonant frequency due to flux noise. The flux noise can perturb the resonant frequency in time, such that the dephasing rate should depend on the derivative, $\gamma_{\phi}\left(\omega_{\rm A}\right) = \eta d\omega_{\rm A} /d\phi_{\rm ex}$, of which the exact formula can be derived from the effective Josephson energy. Here, $\phi_{\rm ex}$ is the flux bias and $\eta$ is a constant to be determined. This interpretation is consistent with our observation, because $d\omega_{\rm A} /d\phi_{\rm ex}$ is increasing with decreasing frequency. By comparison, the external Q factor does not depend on the resonant frequency, which is also consistent with our observation. In these regards, we separate the energy dissipation and dephasing rates from the measured internal loss rate as $\gamma_{i}(\omega_{\rm A}) = \gamma_{0} + \gamma_{\phi}\left(\omega_{\rm A}\right)$. The measured results fit very well with these relations, which indicates a good understanding of the dissipation mechanisms of our system. Besides, we also use a third-order polynomial to fit the weak dependance of the external decay rate on the frequency, $\gamma_{\rm m}\left(\omega_{\rm A}\right) \equiv \gamma_{\rm c}\left(\omega_{\rm A}\right)$, which may originate from possible experimental imperfections. In total, we obtain the total energy dissipation rate $\gamma(\omega_{\rm A})=\gamma_{0} + \gamma_{\rm m}\left(\omega_{\rm A}\right)$ and the dephasing rate $\gamma_{\phi}\left(\omega_{\rm A}\right)$. The characterization result shows that the second resonator is under coupled with $\gamma_{0}=2.26\,{\rm \mu s^{-1}}$ and $\gamma_{\rm m}=1.59\,{\rm \mu s^{-1}}$ on average. The total energy dissipation rate, $\gamma$, dominates the dephasing rate, $\gamma_{\phi}$, for $\omega_{\rm A}/2\pi \geq 6.9\,{\rm GHz}$. This indicates that the dephasing effect may be fairly neglected in this frequency range. We note that $\gamma_{0}$ is a combination of the intrinsic damping rate, $\gamma_{\rm A}$, and the coupling induced damping rate, $\gamma_{\rm f}$, as discussed in Section\,\ref{sec:sample}. With no knowledge on the ratio between the two rates, we simply assume that $\gamma_{\rm A}=\gamma_{\rm f}=\gamma_{0}/2$ in the rest of the discussions, which already shows a good consistency between the simulation and our experimental results.

\subsection{Gain and noise in the output path}\label{sec:output}
We relate the output signal field at the cryogenic temperature, $b_{\rm s,in}$, and the fields to be measured at the room temperature, $b_{\rm s,out}$, by the Caves formula \cite{Caves1982, Mariantoni2010, Renger2021}
\begin{align}
	b_{\rm s, out} \approx \sqrt{G} \left(b_{\rm s,in} + b_{\rm n,in}^{\dagger}\right),
\end{align}
where $b_{\rm n,in}$ is the field operator of the amplification noise, and $G$ is the power gain of the amplification chain. Here, we have neglected the difference between $G$ and $(G-1)$ for a sufficiently large gain ($G \gg 1$), which is valid in common experiments of superconducting quantum circuits. We use the thermal noise as a resource to obtain a precise knowledge of $G$ and $b_{\rm n,in}$ \cite{Menzel2010, Menzel2012, Zhong2013, Fedorov2016, Goetz2017, Fedorov2018, Pogorzalek2019, Fedorov2021}.

The Planck's law describes the energy density of a field emitted by a blackbody thermalized at temperature $T$. A straightforward derivation of Planck's law can be obtained by recalling the properties of a single-mode thermal state at temperature $T$, where the average photon number is $\bar{n}_{T}(\omega) = 1/\left\{\exp\left[\hbar\omega/(k_{\rm B}T)\right]-1 \right\}$. Here, $\bar{n}_{T}(\omega)$ has the dimension of photon number per second per bandwidth. Straightforwardly, the power of thermal radiation in a narrow band, $2B/2\pi$, can be obtained as $P = B\hbar\omega\bar{n}_{T}(\omega)/\pi$. The value of $P$ can be calculated from the measured I/Q quadratures, that is $P=\left(\overline{I^2}+\overline{Q^2}\right)/(2Z_{0})$. Here, we have assumed a perfect impedance match at the ADC input with $Z_0 = 50\Omega$. The factor of $2$ originates from the sinusoidal nature of the microwave field. 

\begin{figure}[hbt]
  \centering
  \includegraphics[width=0.4\columnwidth]{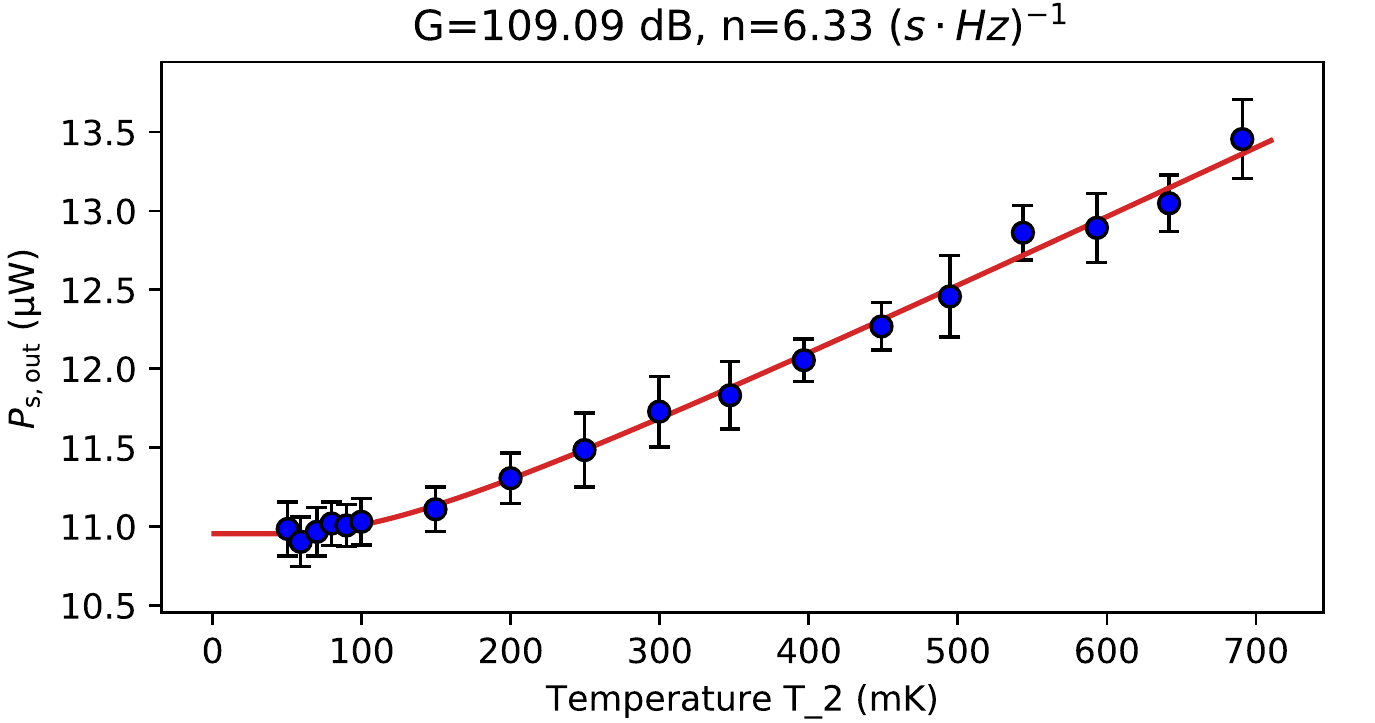}
  \caption{{\bf Experimentally determined parameters of the output channel, OUT\,2.} We tune the second resonator to approximately $7.10\,{\rm GHz}$ and measure the blackbody radiation from the signal path, OUT\,2, within a $\pm 2\,{\rm MHz}$ bandwidth around the central frequency $6.95\,{\rm GHz}$ (blue dots). The blackbody radiation is generated by a $30\,{\rm dB}$ heatable attenuator which is mounted just at the sample input. The error bars represent the standard deviation among $16$ independent experiments, and the red solid curve is the numerical fit.}
  \label{fig:output}
\end{figure}

To characterize the parameters $G$ and $\langle b^{\dagger}_{\rm n,in}b_{\rm n,in}\rangle$, we tightly clamp a homemade cryogenic heater and a homemade cryogenic thermometer to a $30\,{\rm dB}$ attenuator to generate the blackbody radiation at the sample input (see Section\,\ref{sec:cryo_setup} for detail). The heater is a $100\,{\rm \Omega}$ resistor (Vishay MCT 0603), of which the temperature, $T$, is measured and controlled by the AC resistance bridge (Picowatt AVS-47B) and the PID temperature controller (Picowatt TS-530A). The $30\,{\rm dB}$ attenuator can be modeled as a beam splitter which transmits $0.1\%$ of its input signal and $99.9\%$ of the thermal radiation from the environment at temperature $T$. Assuming that the measurement bandwidth is largely detuned from the resonant frequency of the resonator, $\omega_{\rm A}$, this blackbody radiation can be fully reflected at the sample input, and then amplified and measured as a finite power $P_{\rm out}\equiv B\hbar\omega_{\rm A}\langle b_{\rm s, out}^{\dagger}b_{\rm s, out} \rangle/\pi$. In the form of the Caves formula, we have
\begin{align}
	P_{\rm s,out} \approx \frac{GB\hbar\omega_{\rm A}}{\pi} \left[\bar{n}_{T}(\omega_{\rm A}) + n + 1 \right]. \label{eq:output}
\end{align}
Here, we have defined $n\equiv\langle b^{\dagger}_{\rm n,in}b_{\rm n,in}\rangle$, and the constant $1$ comes from the commutation relation, $b^{\dagger}_{\rm n,in}b_{\rm n,in}=b_{\rm n,in}b^{\dagger}_{\rm n,in} - 1$. In our experiment, we calibrate $G$ and $n$ by sweeping the temperature $T$. 

Figure\,\ref{fig:output} shows the relation between the measured power, $P_{\rm s, out}$, and the temperature, $T$, for the output channel OUT\,2. The resonant frequency of the second resonator is tuned to approximately $7.10\,{\rm GHz}$, while we measure the microwave signal at $6.95\,{\rm GHz}$ within a $\pm 2\,{\rm MHz}$ bandwidth. We note that $B/2\pi=2\,{\rm MHz}$ is the cut-off frequency of the low-pass digital filter on FPGA, because the two sidebands of the microwave signal around $6.95\,{\rm GHz}$ are folded into a single sideband during the digital down conversion process. The local temperature of the heatable attenuator is varied from approximately $50$ to $700\,{\rm mK}$ with a precision of $\pm 2.5\,{\rm mK}$ during the measurement time. At each temperature, we average the measured signal power by approximately $5\times 10^4$ times, where the error bars are obtained by repeating this procedure $16$ times. We use the least square estimation method to fit Eq.\,\eqref{eq:output} with the measurement averages. We observe a power gain of $G=109.1\,{\rm dB}$ for the output path, with the mean noise photon number of $n=6.3\,{\rm (s\cdot Hz)^{-1}}$ corresponding to a noise temperature of $2.1\,{\rm K}$. 

\subsection{Attenuation and offset in the input path}\label{sec:input}
\begin{figure}[hbt]
  \centering
  \includegraphics[width=0.9\columnwidth]{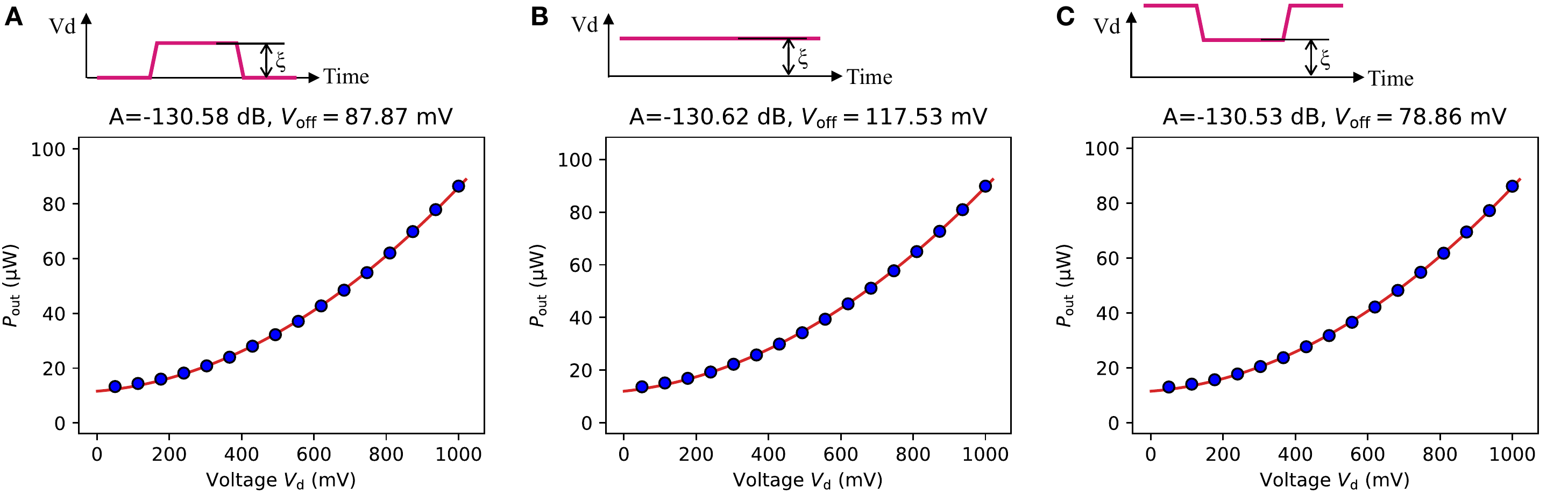}
  \caption{{\bf Experimentally determined parameters of the input channel, MW\,2.} We drive the system through the input path, MW\,2, and measure the reflected signal from the path, OUT\,2, within $\pm 2\,{\rm MHz}$ around the central frequency $7.00\,{\rm GHz}$ (blue dots). The error bars represent the standard deviation among $16$ independent experiments, which is smaller than the size of the dots, and the red solid curve is the numerical fit. Panels (A)-(C) correspond to different pulse shapes.}
  \label{fig:input}
\end{figure}

Having the knowledge of the gain, $G$, and noise photon number, $n$, in the output path, OUT\,2, we move on to characterize the attenuation, $A$, of the input path, MW\,2. Assuming that the power of a signal generator is set as $P_{\rm d}$, we relate the measured signal power, $P_{\rm s,out}$, and $P_{\rm d}$ in a similar form of the Caves formula
\begin{align}
	P_{\rm s, out} \approx G \left[AP_{\rm d} + \frac{B\hbar\omega_{\rm A}}{\pi}\left(n+1\right)\right]. \label{eq:input}
\end{align}
The aim of determining $A$ is to establish a relation between $P_{\rm d}$ and the driving strength, $\xi$, in the system Hamiltonian, that is \cite{Ong2011}
\begin{align}
	\xi = -i\sqrt{\gamma_{\rm m/f} AP_{\rm d}/\hbar \omega_{\rm d}}.
\end{align}
Here, we have assumed the driving field to be in a coherent state. In our experiment, we drive the system via a homemade pulse-shaping module, where the carrier wave generated by the signal generator (R\&S SMF100A) is modulated by a voltage signal, $V_{\rm d}$, generated by the AFG (Tektronix AFG3252), as described in Section\,\ref{sec:rt_setup}. We assume a simple relation between $P_{\rm d}$ and the pulse amplitude $V_{\rm d}$ as $P_{\rm d} = \left( V_{\rm d} - V_{\rm off}\right)^2/(2Z_{0})$. Here, $V_{\rm off}$ is the offset voltage in the setup, which originates from the imperfect grounding of mixers in the pulse-shaping module. Besides, we assume a perfect impedance match with $Z_{0}=50\,{\rm \Omega}$. The goal of the input characterization experiment is to determine the values of $A$ and $V_{\rm off}$.

Figure\,\ref{fig:input} shows the measured signal power, $P_{\rm out}$, as a function of the pulse amplitude, $V_{\rm d}$, for the input channel, MW\,2. Here, we set the carrier frequency of the input field to $7.00\,{\rm GHz}$ and vary the pulse amplitude from $50\,{\rm mV}$ to $1000\,{\rm mV}$. The other parameters are set to be exactly the same as for the output characterization experiments. We employ three different pulse shapes for characterization. Correspondingly, the characterized attenuations are $A=-130.6\,{\rm dB}$, $-130.6\,{\rm dB}$, and $-130.5\,{\rm dB}$, respectively, which are almost identical to each other. However, the offset voltage shows a clear dependance on the pulse shape. The results are $V_{\rm off}=88\,{\rm mV}$, $118\,{\rm mV}$, and $79\,{\rm mV}$, which vary by approximately $50\,{\rm mV}$ for the three different pulse shapes shown in Fig.\,\ref{fig:input}. Besides the imperfect grounding, this may also be attributed to the finite on/off ratio of the mixer, which mixes the carrier wave with the voltage signal in the pulse-shaping module. However, we note that a $50\,{\rm mV}$ offset voltage corresponds to an inaccuracy of $\xi/2\pi$ being less than $5\,{\rm kHz}$, which is negligibly small in all of our experiments.

\section{Experimental methods}\label{sec:methods}
\begin{figure}[hbt]
  \centering
  \includegraphics[width=0.6\columnwidth]{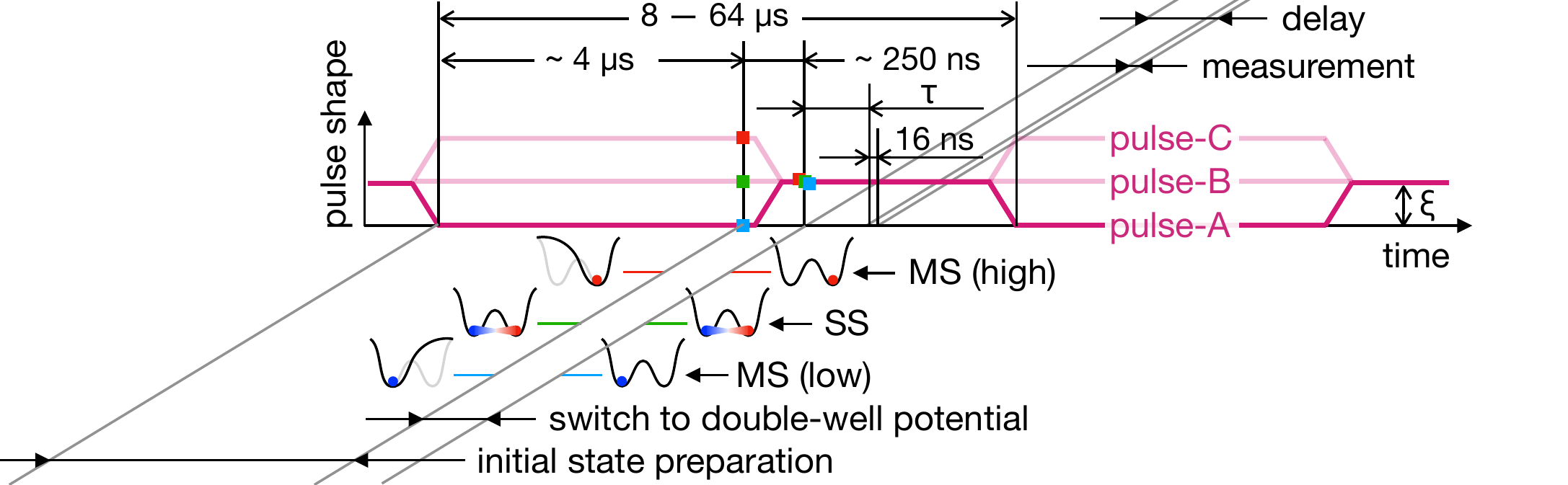}
  \caption{{\bf The schematic of the pulsed heterodyne measurement protocol .} We prepare the initial state of system in one of the two potential wells by driving it with either a \textit{zero}-amplitude (pulse-A) or a high-intensity field (pulse-C). Then, the driving strength is switched to $\xi$ and lasts for a controllable time $\tau$ before we perform a $16\,{\rm ns}$ quadrature measurement. This procedure is repeated for more than $10^{6}$ times to accumulate a histogram of the field quadratures. In certain experiments, we also drive the system with a constant driving field with driving strength $\xi$ (pulse-B), which prepares the system in the SS at the initial time.}
  \label{fig:methods}
\end{figure}

The pulsed measurement plays a fundamental role in revealing the non-equilibrium quantum dynamics of the Duffing oscillator, as is schematically shown in Fig.\,\ref{fig:methods}. The basic logic of the experiment is (i) to prepare the system in one of the two wells, (ii) to drive the system at $\xi$, (iii) to wait for a controllable time $\tau$, and (iv) to start a short measurement. We note that the control parameters are not swept in a continuous fashion, which is different from the relevant experiments in the literature. 

\subsection{Initial state preparation}\label{sec:pulse}
To prepare the system in different wells at the initial time, we set the driving strength at either \textit{zero} or the maximum value one can achieve (pulse-A and C), which is limited by approximately $4.7\,{\rm V}$ at the AFG output when using the pulse-generation mode. As discussed in Section\,\ref{sec:duffing_classical}, the system has a single-well potential at a sufficiently small or large driving strength, which corresponds, respectively, to one of the two wells. We wait for approximately $4\,{\rm \mu s}$ to let the system reach the SS of the single-well potential, which is more than $10$ times larger than the free relaxation time, $1/\gamma$. Next, we switch the driving strength to $\xi$, which defines the driving strength in Eqs.\,\eqref{eq:duffing_langevin} and \eqref{eq:duffing_classical}. The switching time is usually set to $250\,{\rm ns}$, which aims to provide a smooth but relatively fast transition between the initial and final values. Depending on the exact value of $\xi$ and $\Delta$, the system can have a double-well potential in the so-called hysteresis regime, while the initial state is prepared in either of the two wells. We also drive the system with a constant driving field (pulse-B), where the system is initially prepared in the SS. 

\subsection{Pulsed measurement}
We always wait for a time duration of $\tau$ before starting a measurement. In order to capture the non-equilibrium dynamics of the system, we measure only one period of the IF signal, which is $16\,{\rm ns}$. Here, only $4$ data points are recorded in a single measurement event, corresponding to one data point of the field quadratures, ${I}+i{Q}$, with a time resolution of $16\,{\rm ns}$. Then, we initialize the system and repeat the same measurement procedure by $10^{6}$ -- $10^{9}$ times depending on the required measurement accuracy, each of which is triggered at the same relatively time after the initialization. Because the experimental conditions are kept the same, the measurement results should also be the same within the uncertainty range defined by the quantum fluctuations. One can figuratively understand the pulsed measurement as using millions of ADCs that measure the system at the same time. Then, we concatenate the data recorded by the different virtual ADCs into a long trace and apply a low-pass filter to increase the SNR. Because the resulting signal is not sequenced in real time, the cut-off frequency, or the ring-up time, of the filter does not influence the time resolution of the measurement result. It indicates that one can apply a relatively narrow-band digital low-pass filter ($\sim 2\,{\rm MHz}$ in this case) to improve the SNR but keep the $16\,{\rm ns}$ time resolution of the result. 

\subsection{Photon correlations}
During the pulsed measurement, we record the two signal moments, $\langle b_{\rm s, out} \rangle$ and $\langle b_{\rm s, out}^{\dagger}b_{\rm s, out} \rangle$, as well as the histogram of $b_{\rm s, out}$ in a $128\times 128$-dimentional matrix. The measured histogram is the Q function of the output field, $b_{\rm s,out}$, which is a convolution between the input field, $b_{\rm s,in}$, and the noise field, $b_{\rm n,in}$, \cite{Kim1995, Kim1997, Eichler2012}
\begin{align}
    Q_{\rm s,out}(\gamma,\gamma^*) &= \frac{1}{G-1}\int d\alpha^2 Q_{\rm s,in}(\alpha,\alpha^*)
	P_{\rm n,in}\left(\frac{\gamma^* - \sqrt{G}\alpha^*}{\sqrt{G-1}},\frac{\gamma - \sqrt{G}\alpha}{\sqrt{G-1}}\right), \label{eq:Q}
\end{align}
where $Q_{\rm s,out}$, $Q_{\rm s,in}$, and $P_{\rm n,in}$ are the quasi-distribution functions of the three fields. Combining Eq.\eqref{eq:Q} with the input-output relation, $b_{\rm s,in}=\sqrt{\gamma_{\rm m}}a$, one can calculate all orders of the signal moments as
\begin{align}
	\langle b_{\rm s, out}^{\dagger k} b_{\rm s, out}^{l} \rangle 
	&= \left(\frac{\gamma_{\rm m}BG}{\pi}\right)^{\frac{k'+l'}{2}}\sum_{k'=0}^{k}\sum_{l'=0}^{l}
	C_{k}^{k'}C_{l}^{l'}
	\langle a^{\dagger k'}a^{l'} \rangle
	\langle b_{\rm r, out}^{\dagger k-k'}b_{\rm r, out}^{l-l'} \rangle,
\end{align}
where $\langle b_{\rm r, out}^{\dagger k} b_{\rm r, out}^{l} \rangle 
	= \left(BG/\pi\right)^{\frac{k+l}{2}}\langle b_{\rm n, in}^{k}b_{\rm n, in}^{\dagger l}\rangle$ is measured when the resonator is in the vacuum state, $C_{k}^{k'}$ is the binomial coefficient. The filter bandwidth is typically set to $B/2\pi=2\,{\rm MHz}$. However, because the coupling strength between the resonator and the microwave line, MW\,2, is approximately $2$ times smaller than $2B$ (see Section\,\ref{sec:open}), we compensate the power gain in our analysis code by $-3\,{\rm dB}$ to characterize the intra-resonator photon number. This correction leads to a good agreement between all of our experimental results and the simulations with no fitting parameter. In certain tasks where a high-precision estimation of the photon number is required, for example, the quantum state tomography discussed in Section\,\ref{sec:tomography}, we will fine tune the value of $G$ in a $\pm 1\,{\rm dB}$ range.

\section{Supplementary data}
\subsection{Closure of the hysteresis loop in the long-time limit}
\begin{figure}[hbt]
  \centering
  \includegraphics[width=0.6\columnwidth]{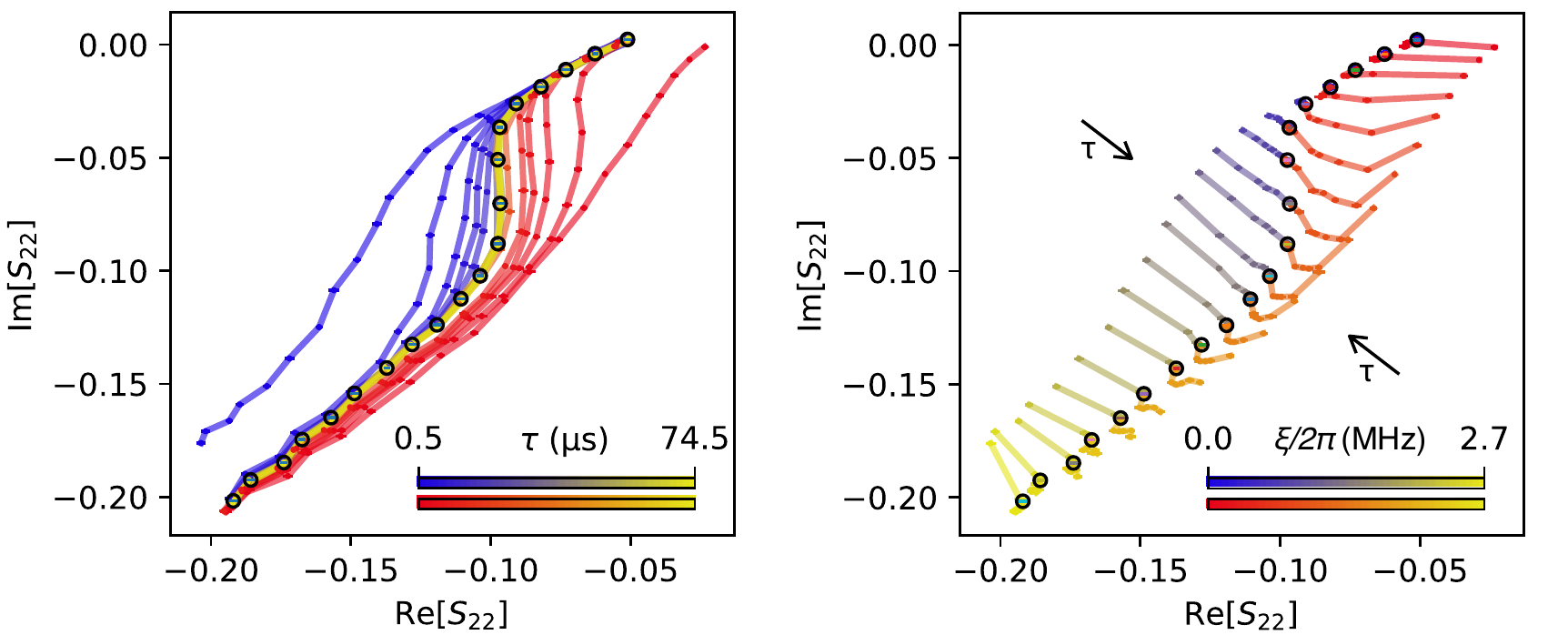}
  \caption{{\bf Power-delay sweep obtained by a reflection-type measurement.} The reflection coefficients, $S_{22}$, corresponding to the two MSs branches (blue and red) form a closed loop, which converge to the unique SS solution (back circle) with increasing $\tau$. Here, the error bars represent the standard deviation over $8$ independent experiments.}
  \label{fig:loop}
\end{figure}

One major difference between the classical and quantum theories of the Duffing oscillator is the number of SSs. The former predicts two in the hysteresis regime, which are localized in either of the two potential wells. However, the latter predicts one unique SS in the entire parameter space. In this regard, one straightforward way to verify the quantum dynamics of the Duffing oscillator is to prepare the system in either of the two wells and wait for a long time before measurement. In the absence of thermal noise, the two classical SSs remain in the well such that the area of the hysteresis loop should not decrease with $\tau$. However, the loop area must decrease in the quantum perspective, because of the uniqueness of the SS. This latter prediction is confirmed in Fig.\,2A of the main text. Moreover, the two branches must converge to a single curve corresponding to that of the SS when $\tau \gg 1/\min_{\xi}\delta_{1}(\xi)$, where $\delta_{1}(\xi)$ is the Liouvillian gap as a function of the driving strength, $\xi$. This phenomenon is not demonstrated in Fig.\,2A of the main text, because $\tau$ is limited by $45\,{\rm \mu s}$ there and we did not measure the SS curve in that experiment. As a supplementary data, we plot in Fig.\,\ref{fig:loop} a similar measurement where $\tau$ goes up to $75\,{\rm \mu s}$. In this case, the SS is also measured. Here, the closure of the hysteresis loop is observed at $\tau \simeq 55\,{\rm \mu s}$, which is larger than $230$ times of the free relaxation time $1/\gamma$. In addition, the two MS branches converge continuously to the SS curve. This result, in combination with Fig.\,2A of the main text, demonstrates the uniqueness of the SS, as predicted in the quantum theory.

\subsection{Extracting the Liouvillian gap from time-domain measurements}
\begin{figure}[hbt]
  \centering
  \includegraphics[width=0.6\columnwidth]{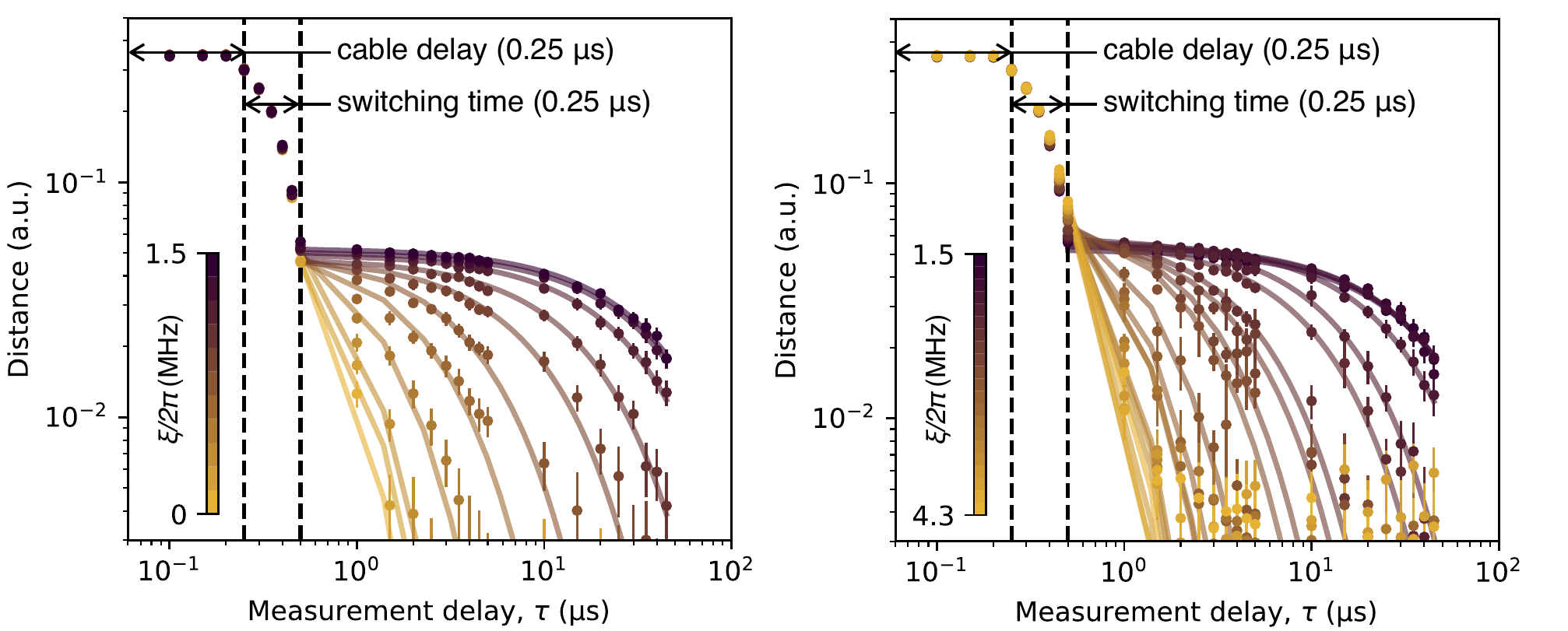}
  \caption{{\bf Power-delay sweep obtained by a reflection-type measurement.} Shown are the raw data (dots) for extracting the Liouvillian gap, and the exponential fitting results (solid curves). With the increase of the driving strength in the $0\leq \xi/2\pi \leq 1.5\,{\rm MHz}$ regime, the relaxation process becomes increasingly slower. However, the relaxation becomes increasingly faster if we increase further the driving strength, $1.5\,{\rm MHz}\leq \xi/2\pi \leq 4.3\,{\rm MHz}$. In all the panels, the error bars represent the standard deviation over $16$ independent experiments.}
  \label{fig:gap}
\end{figure}

To extract the Liouvillian gap, $\delta_{1}$, from the time-domain measurements, we calculate the distance between the two MSs branches for each $\xi$ as a function of $\tau$. Fig.\,\ref{fig:gap} shows the raw data of Fig.\,2C of the main text with fitted results. Here, we did not correct the cable delay in the time axis, which is measured to be approximately $250\,{\rm ns}$. For each $\xi$, we fit the data in the $\tau \geq 0.5\,{\rm \mu s}$ range with an exponential function. Because $\tau > 1/\gamma$, the fitted decay rate can be fairly regarded as the Liouvillian gap, which dominates the relaxation of the system in the long-time limit. As described in the main text, the fitted value of $\delta_{1}$ is approximately equal to the energy dissipation rate, $\gamma$, at either low or high driving strengths. However, it decreases over two orders of magnitude when approaching to the critical driving strength, $\xi^{*}/2\pi=1.51\,{\rm MHz}$. This result explains the two-stage relaxation process of the system, as shown in Fig.\,2B of the main text.

\subsection{Squeezing levels in the two phases besides the phase transition}\label{sec:phase}
\begin{figure}[hbt]
  \centering
  \includegraphics[width=0.5\columnwidth]{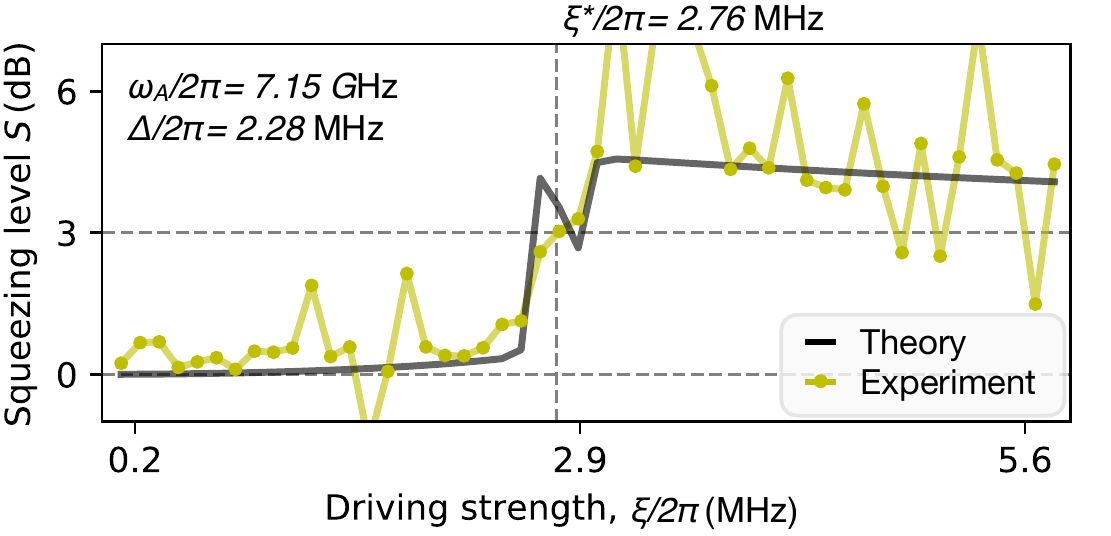}
  \caption{{\bf The squeezing level as a function of the driving strength.} The critical point, $\xi^{*}/2\pi=2.76\,{\rm MHz}$, separates the system into two different phases with drastically different squeezing levels $S$. The value of $S$ is approximately \textit{zero} before the transition, which indicates a coherent phase of the system. After the transition, the squeezing level is approximately $3\,{\rm dB}$, which corresponds to a squeezed phase. Here, we fine tune the estimated resonant frequency to $\omega_{\rm A}/2\pi=7.15\,{\rm GHz}$, and the detuning frequency is $\Delta/2\pi=2.28\,{\rm MHz}$.}
  \label{fig:squeezing}
\end{figure} 

Because the SS in the two phases is approximately either a coherent or squeezed state, as shown in Fig.\,4 of the main text, we use a Gaussian function to describe them and calculate the corresponding squeezing levels \cite{Bajer2004}. By definition, a Gaussian state is a rotated, squeezed, and displaced thermal state $\rho = D(\alpha) S(\zeta) R(\phi) \rho_{T} R^{\dagger}(\phi) S^{\dagger}(\zeta) D^{\dagger}(\alpha)$, where $D(\alpha)$, $S(\zeta)$, and $R(\phi)$ are the displacement, squeeze, and rotation operators. The squeezing level can be defined as $S = -20\left|\zeta\right|\log_{10}(e)$ where $e$ is the exponential constant. On the other hand, for Gaussian states we have 
\begin{align}
	\tanh\left(2\left|\zeta\right|\right) = \frac{\langle a^2 \rangle - \langle a \rangle^2}{\langle a^{\dagger}a \rangle + 1/2 - \left|\langle a \rangle\right|^2}.
\end{align}
One can thus calculate the squeezing level of the two phases according to the measured signal moments: $\langle a \rangle$, $\langle a^{\dagger}a \rangle$, and $\langle a^2 \rangle$. Fig.\,\ref{fig:squeezing} shows the squeezing level of the system as a function of the driving strength. The critical point, $\xi^{*}/2\pi=2.76\,{\rm MHz}$, separates the system into two different phases with drastically different squeezing levels. The value of $S$ is approximately \textit{zero} before the phase transition, but jumps to approximately $3\,{\rm dB}$ afterwards. This observation reveals the two distinct phases of the DPT with respect to the different squeezing levels. We note that the Gaussian-state approximation breaks down around the critical point, where the SS is a mixture of the two phases \cite{Bajer2004}. This can be seen from the unexpected wiggle in the theory curve around $\xi^{*}$.

\subsection{Quantum state tomography of the phase transition process}\label{sec:tomography}
\begin{figure}[hbt]
  \centering
  \includegraphics[width=\columnwidth]{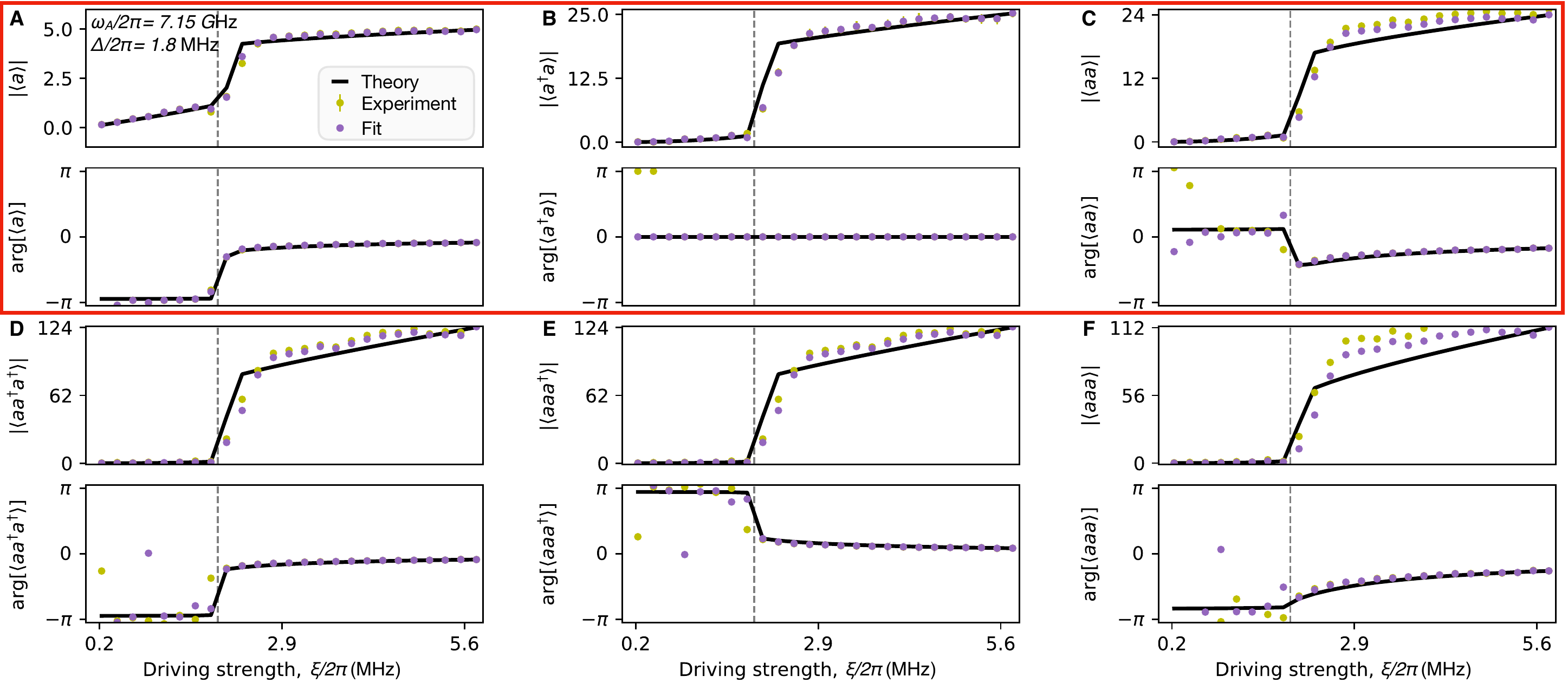}
  \caption{{\bf The first three orders of signal moments.} The measured amplitude and phase of the signal moments show an excellent fit to the theoretical prediction in Eq.\,\eqref{eq:positive_p} with no fitting parameter, which provides the opportunity for quantum state tomography. {\bf (A)}-{\bf (F)} correspond to the signal moments, $\langle a \rangle$, $\langle a^{\dagger}a \rangle$, $\langle a^2 \rangle$, $\langle aa^{\dagger 2} \rangle$, $\langle a^{2}a^{\dagger} \rangle$, $\langle a^3 \rangle$, respectively. The error bars in (A) and (B) represent the standard deviation over $8$ independent experiments, while it is not recorded in other panels. The first two orders of moments, enclosed by the red box, are used for quantum state tomography.}
  \label{fig:moments}
\end{figure}

Quantum state tomography in our experiment is achieved by combining Eqs.\,\eqref{eq:positive_p} and \eqref{eq:exact_solution}. On the one hand, the exact Wigner function can be fully determined by the two parameters, $c$ and $d$, in Eq.\,\eqref{eq:exact_solution}. On the other hand, these two parameters are closely related to the signal moments in Eq.\,\eqref{eq:positive_p}. We thus find the best fit of $c$ and $d$ from the first two orders of signal moments according to Eq.\,\eqref{eq:positive_p}, and insert the fitted values to Eq.\,\eqref{eq:exact_solution} to get the Wigner function. Because $c$ is a complex number and $d$ is real, one needs at least the information of the two moments, $\langle a \rangle$ and $\langle a^{\dagger}a \rangle$, to determine the two parameters. Here, we take also the $\langle a^2 \rangle$ term into consideration, which makes the fitting problem overdetermined, and thus increases the reliability of the tomography result.

To minimize the influence of the dephasing effect, we only perform quantum state tomography at $\omega_{\rm A}/2\pi=7.15\,{\rm GHz}$. Fig.\,\ref{fig:moments}A-F compares the measured signal moments and the simulation results up to the third order, which shows a good  agreement between theory with no fitting parameter and experiment. Here, we fine tune the power gain by a factor of $-0.57\,{\rm dB}$ for all the orders of signal moments according to the last data point. We also adjust the global phase offset of each signal moments according to the last data point. The shown excellent fit between theory and experiment thus justifies the feasibility of extracting the values of $c$ and $d$ from the measured signal moments for quantum state tomography.  The tomography result, as shown in Fig.\,4 of the main text, is obtained from the first $2$ orders of signal moments shown in Fig.\,\ref{fig:moments}A-C.

We comment that the described procedure relies on the priori knowledge of the SS and thus lacks general objectivity. Alternative methods, such as coupling a probe qubit to the resonator \cite{Shalibo2013, Kirchmair2013}, may provide a more general tomography result. However, we note that the model itself does not necessarily predict the non-classical feature of the SS in the $0.52\,{\rm MHz} \leq \xi_{0}/2\pi \leq 0.64\,{\rm MHz}$ range, as shown in Fig.\,4 of the main text. We see that we are able to reveal the transition process with a simple physical model but no fitting parameter, and also to obtain a consistent understanding between theory and experiment among independent experiments.

\subsection{The dephasing effect and possible two-photon processes}\label{sec:dephasing}
\begin{figure}[hbt]
  \centering
  \includegraphics[width=\columnwidth]{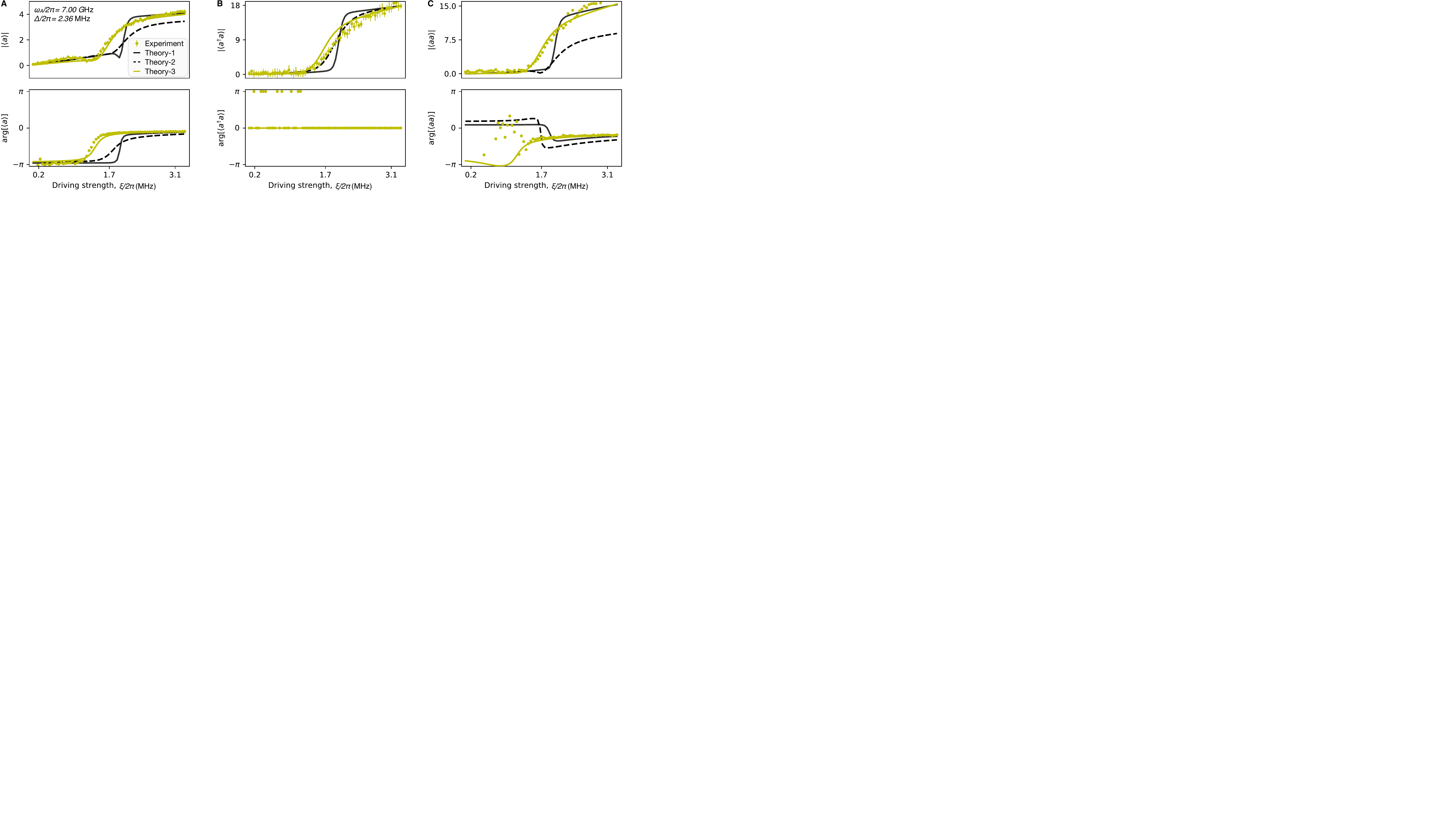}
  \caption{{\bf Comparison between experimental and numerical results for the first two orders of signal moments.} The yellow dots represent the experimental results, where the error bars represent the standard deviation over $8$ independent experiments. The black solid curves show the analytical result in Eq.\,\eqref{eq:positive_p}, where only energy dissipation is considered (Theory-$1$). The dashed black line shows the master equation simulation with dephasing effect and a finite thermal photon number of the environment (Theory-$2$). It captures the the slower transition rate observed in experiment but leads to a large discrepancy in $\langle a^2 \rangle$. We also consider a weak two-photon drive and loss process in the model (Theory-$3$, yellow solid), which provides a good agreement with all the three signal moments.}
  \label{fig:dephasing}
\end{figure}

So far, we have neglected dephasing effects in the discussion. This is feasible because the determined dephasing rate, $\gamma_{\phi}$, is smaller than the energy dissipation rate, $\gamma$, in the frequency range of interest (see Section\,\ref{sec:open} for the characterization results). It is also justified by showing the excellent agreement between theory and experiment for all the first three orders of signal moments, as shown in Fig.\,\ref{fig:moments}. However, this agreement exists only at high resonant frequencies. As can be seen in Fig.\,3 of the main text, the experimentally observed transition curve is less steep than that predicted by the model at lower frequencies. As discussed in Section\,\ref{sec:sample}, we attribute the reduced steepness to the presence of finite dephasing, since the dephasing rate increases when going to lower frequencies. 

To achieve a quantitive understanding of the experimental data, we add the dephasing term by hand and move further to the Schr{\"o}dinger picture. The master equation in the Lindblad form reads
\begin{align}
	\partial_t \rho(t) = -i\left[ H_{\rm eff}, \rho(t) \right] 
	+ \frac{\gamma}{2}\left(n_{\rm T}+1\right)\mathcal{D}\left[a\right]\rho(t) 
	+ \frac{\gamma}{2}n_{\rm T}\mathcal{D}\left[a\right]\rho(t) 
	+ \frac{\gamma_{\phi}}{2}\mathcal{D}\left[a^{\dagger}a\right]\rho(t). \label{eq:duffing_langevin_dephasing}
\end{align}
Here, $\rho(t)$ is the density operator, $H_{\rm eff}$ is the effective Hamiltonian of the system, and $\mathcal{D}\left[a\right]$ and $\mathcal{D}\left[a^{\dagger}a\right]$ are the Lindbladian superoperators. Besides, we consider also a finite temperature of the bath $n_{T}$. The value of the energy relaxation and the dephasing rates, $\gamma$ and $\gamma_{\phi}$, have been determined in Section\,\ref{sec:open}. 

Figure\,\ref{fig:dephasing} compares the measured signal moments with the simulation results. Compared with the analytical result with $\gamma_{\phi} = 0$, a finite dephasing rate, $\gamma_{\phi}$, nicely captures the observed smaller steepness of the transition. Here, we have also assumed a small thermal photon number of the environment, $\bar{n}_{T}=0.1$. However, a closer inspection of the second-order moment, $|\langle a^2 \rangle|$, indicates that $\gamma_{\phi}$ also leads to a significantly smaller saturation value of this quantity. To achieve a better fitting between the simulation and the experiments, one may consider to include the second-order processes into the simulation, which has been neglected for deriving Eq.\,\eqref{eq:langevin}. Here, we consider the two-photon drive, $\xi_{2}\left(a^2+a^{\dagger 2}\right)$, and correspondingly the two-photon loss, $\left(\gamma_{2}/2\right)D\left[a^{2}\right]$. These higher-order processes should be weak, such that the parameters, $\xi_{2}$ and $\gamma_{2}$, are assumed to be smaller than $\xi$ and $\gamma$, respectively. We achieve a quantitive agreement between theory and experiment for $\xi_{2}=0.3\xi$ and $\gamma_{2}=0.1\gamma$. These results demonstrate that we are able to achieve a consistent interpretation of our experimental results within a simple physical model. Nevertheless, we emphasize that the conclusions drawn from our experiment are either insensitive to the dephasing rate, such as the hysteretic behavior (Fig.\,1 of the main text), two-stage relaxation process (Fig.\,2 of the main text), or based on the high-frequency measurements where the dephasing rate is much smaller than the energy dissipation, such as the increasingly sharp transition step with scaling factor $N$ (Fig.\,3 of the main text) and the quantum state tomography results (Fig.\,4 of the main text). 

\bibliography{Duffing_REF}